# Thermophotovoltaic performance metrics and techno-economics: efficiency vs. power density

Shomik Verma, Kyle Buznitsky, Asegun Henry*

Department of Mechanical Engineering, Massachusetts Institute of Technology,
77 Massachusetts Avenue, Cambridge, MA 02139, U.S.A.
*ase@mit.edu

## Abstract

Thermophotovoltaics (TPV) are a promising new approach for converting heat to electricity. Their performance is primarily characterized by two metrics: efficiency and power density. While recent works have shown high efficiency, it is important to understand how both of these metrics impact the techno-economics of a TPV system as efforts to commercialize the technology advance. In this work, we develop the first unification of efficiency and power density into a single techno-economic metric based on the levelized cost of electricity (LCOE). We find that the LCOE can be broken into two parts: heating cost, including infrastructure and inputs for providing heat to the TPV cells, and cell cost, the capital cost of the TPV cells. We show that systems with high heating costs should prioritize TPV efficiency, while systems with high cell costs should prioritize power density. We then develop a model to identify the most impactful cell properties in improving the important performance metric and reducing system LCOE. Namely, improving spectral control with increased back-surface reflectance is the most effective to reduce LCOE in systems with high infrastructural costs, while increasing the view factor and reducing front-surface reflectance are most critical in systems with high TPV cell cost. Improving just one or two of these properties can reduce the LCOE by 25-75%, reaching competitive values ~ 8 ¢/kWh-e, less than the average cost of electricity in the US. This study thus elucidates which TPV performance metric is more important for system technoeconomics and how to maximize it.

# Nomenclature

*Abbreviations*

| | |
|---|---|
| ABE | Above-bandgap emissivity |
| ABR | Above-bandgap reflectance (front-surface) |
| BG | Bandgap (eV) |
| BW | Bandwidth |
| CAPEX | Capital expenditure ($) |
| CF | Capacity factor |
| CPA | Cell cost per area ($/cm$^2$) |
| CPE | Cost per energy ($/Wh) |
| CRF | Capital recovery factor |
| LCOE | Levelized cost of electricity ($/Wh-e) |
| LCOH | Levelized cost of heating ($/Wh-th) |
| NRR | Non-radiative recombination ratio |
| OPEX | Operational expenditure ($) |
| SBR | Sub-bandgap reflectance (back surface) |
| TPV | Thermophotovoltaics |
| VF | View factor |

*Variables*

| | |
|---|---|
| $A_{TPV}$ | TPV cell area (cm$^2$) |
| $CAPEX_{infras}$ | Infrastructure capital expenses ($) |
| $CPE_{th,infras}$ | Infrastructure cost per thermal energy output ($/Wh-th) |
| $CPE_{th,input}$ | Input cost per thermal energy output ($/Wh-th) |
| $i$ | Interest rate |
| $I$ | Current (A) |
| $n$ | System lifetime (years) |
| $P_{dens,TPV}$ | TPV power density (W/cm$^2$) |
| $P_{elec}$ | TPV electrical power produced (W) |
| $P_{in}$ | Input power (W) |
| $P_{rad}$ | Net heat transfer to TPV cells (W) |
| $Q_{abs}$ | TPV heat absorbed/generated (W) |
| $Q_{loss,heat}$ | Losses in heat device (W) |
| $Q_{loss,TPV}$ | Losses in TPV cell (W) |
| $R_{series}$ | Series resistance (Ω cm$^2$) |
| $T_{emit}$ | Emitter temperature (°C) |
| $t_{out}$ | Time per year the TPV produces electricity (h) |
| $V_{oc}$ | Open-circuit voltage (V) |

*Greek symbols*

| | |
|---|---|
| $\eta_{rad}$ | Radiative efficiency |
| $\eta_{TPV}$ | TPV cell efficiency |
| Ω | Resistance (Ohms) |

# 1. Introduction

Thermophotovoltaics (TPV) are rapidly gaining traction as an alternative means of converting heat to electricity for power applications, as compared to the predominant state of the art, e.g., turbines [1]. Recent efficiency gains in TPV, notably surpassing the average turbine (~ 35%) [1,2], and high-level techno-economic analyses showing competitive cost and performance [3,4] have generated a new wave of interest in the technology.

TPV works by converting the light emitted radiatively from a heat source into electricity, using the same physics (i.e., the photovoltaic (PV) effect) as solar cells, but utilizing a terrestrial heat source instead of the sun [5]. TPV has many benefits including being solid state with no moving parts and having a higher theoretical efficiency than many heat engines [6,7]. TPV cell efficiency differs from solar PV, though both are defined as electricity output over power input. For solar PV, the input is the rate at which light reaches the cell, with any reflected light lost to space due to the sun's distance and low view factor. In contrast, TPV uses a terrestrial heat source with a high view factor to the TPV cell, allowing reflected light from the TPV cell to return to the emitter, preserving energy and, importantly, keeping the emitter hot. Thus, TPV cell efficiency is defined as the ratio of electric power output to total absorbed power [7–10]:

$$\eta_{TPV} = \frac{P_{elec}}{P_{elec} + Q_{abs}}$$

where $P_{elec}$ is electric power output and $Q_{abs}$ is waste heat absorbed/generated by the cell. Since the efficiency determines the amount of output power for a given input power, low efficiency cells can result in large amounts of wasted energy input, increasing costs.

Recent works have demonstrated high-efficiency TPV cells (>40%) [1,2,11]. However, the efficiency only tells one side of the story. As a PV device, TPV operates by converting photons with energy above its bandgap into electricity. Hypothetically, one can imagine a device that filters out all wavelengths of light except the photons at the bandgap of the cell, theoretically approaching the Carnot efficiency (e.g. 87% with an emitter temperature of 2000°C) [6]. However, the electricity produced from such a device would be vanishingly small, as the number of photons right at the bandgap is low. Similarly, a TPV device with a high bandgap would have low thermalization losses since the above-bandgap photons would be closer in energy to the bandgap, but due to Planck's law the number of photons above the bandgap would be low. Thus, another important metric for TPV cells is power density, defined as the electricity produced divided by the device area [8,12,13]:

$$P_{dens,TPV} = \frac{P_{elec}}{A_{TPV}}$$

Low power density cells can result in a large-area system, and since the cost of a TPV or PV cell scales with its area, this increases costs.

Both of these performance metrics are therefore important for TPV and are reported in previous works [5,14–16]. Ideally, we would want to maximize both efficiency and power density for the best device. However, as discussed above, these metrics may be in conflict. As efforts to commercialize TPV develop, it is important to understand the trade-offs between efficiency and power density, specifically how these two TPV performance metrics impact the techno-economics of the overall system.

Previous works have developed unified metrics including both efficiency and power density, but these have been based on only the performance of the TPV cell instead of the system as whole, likely in an effort to isolate optimization of cells [17,18]. Previous works have also discussed TPV techno-economics but have been restricted to only a few specific systems, as summarized in Table 1.

*Table 1: List of existing research evaluating techno-economics of TPV systems.*

| System application | Metric | Value | Ref |
|---|---|---|---|
| Residential heating/cooling | Levelized cost of electricity | 12.8 ¢/kWh | [4] |
| Residential heat/electricity | Levelized cost of combined heat and power | 10 ¢/kWh | [19] |
| Residential heat/electricity | Yearly money savings | 1.8-3.9 €/m$^2$ | [20] |
| Residential electricity | Levelized cost of electricity | 6-25 €¢/kWh | [21] |
| Residential electricity | Levelized cost of electricity | 5-11 €¢/kWh | [22] |
| Latent heat storage | Levelized cost of electricity | 14.5-16.5 €¢/kWh | [23] |
| Sensible heat storage | System (capital) cost | 50 $/kWh-e | [3,24] |

A more comprehensive analysis for TPV techno-economics explicitly evaluating the impacts of efficiency and power density has been lacking. Therefore, in this work we seek to answer the following questions: how do each of the performance metrics impact the techno-economics of a TPV system? Which performance metric is more important for a given system? Given its importance, how can we design a TPV cell (bandgap, cell properties) to maximize techno-economic viability?

## 2. Methods

### 2.1. LCOE unifies the performance metrics

To understand when each metric is more important, consider a TPV system (shown in Figure 1) that takes a power input $P_{in}$ (fuel, electricity, sunlight) and through some system infrastructure (resistance heaters, tubing, emitters) converts that power into radiated (net) power to the TPV cells $P_{rad} = P_{in}\eta_{rad}$. The radiative efficiency $\eta_{rad}$ includes inefficiencies in converting the input power to heat, heat loss from the device due to conduction and convection, and any radiation from the emitter lost to the environment – further details are provided in Section S1 and examples of specific systems are provided in Section S2. Then, the TPV efficiency impacts how much electricity can be output ( $P_{elec} =$ $P_{in}\eta_{rad}\eta_{TPV}$), and the power density impacts how much TPV area is required to output that power $\left(A_{TPV} = \frac{P_{elec}}{P_{dens,TPV}}\right)$.

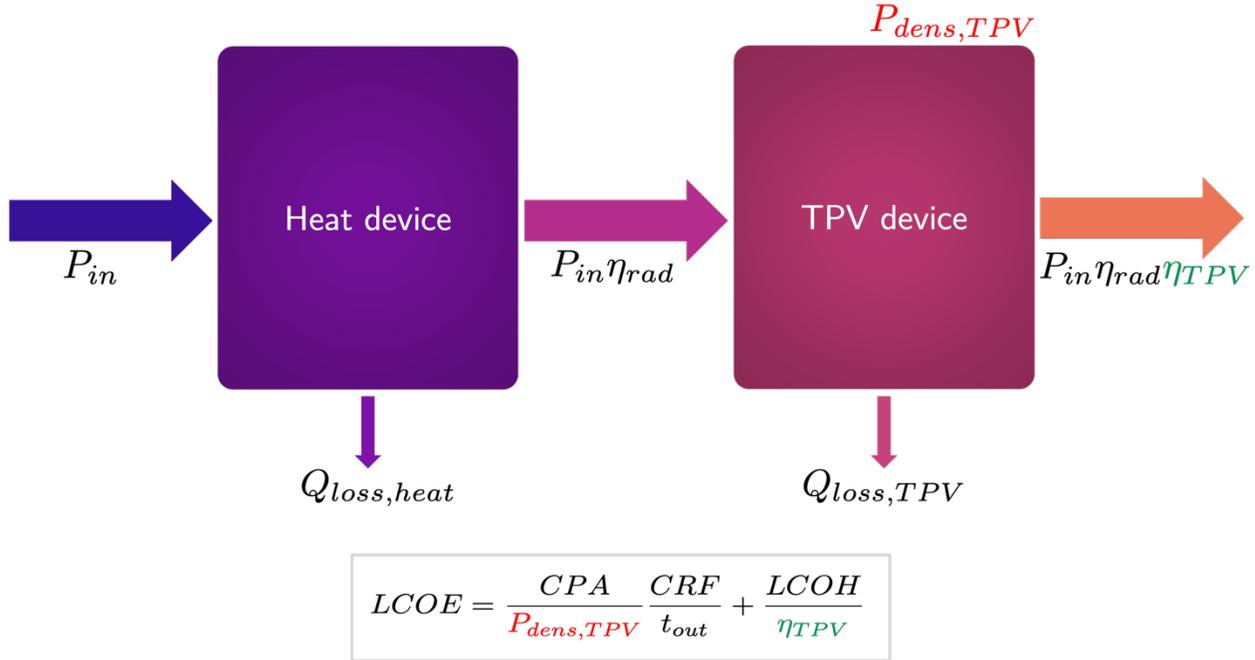

Figure 1: Schematic of a general TPV system. Power input $P_{in}$, heat device to convert that power into light hitting the TPV cells $P_{rad} = P_{in}\eta_{rad}$ at radiative efficiency $\eta_{rad}$, and TPV which converts light to electricity $P_{elec} = P_{rad}\eta_{TPV}$ at TPV efficiency $\eta_{TPV}$. Also shown is the key LCOE metric developed in this work, where CPA is the cell cost per area, CRF is the capital recovery factor, $t_{out}$ is the time the TPV outputs power, and LCOH is the levelized cost of heating defined as the amortized heat device cost plus the fuel cost.

We notice that both efficiency and power density impact the cost – low efficiency means more power input (e.g. fuel) is required, and low power density means more TPV area is required, to generate the same output power. To evaluate the cost more rigorously, we can use the levelized cost of electricity (LCOE) commonly used for power generation technologies. Here we use the simplified form of LCOE [25] which assumes constant values of fuel expenditures, energy generation, and interest rates year-over-year, to allow fair comparison between different scenarios. For a general TPV system as shown in **Figure 1**, the LCOE expression becomes

$$LCOE = \frac{CPA}{P_{dens,TPV}} \frac{CRF}{t_{out}} + \frac{LCOH}{\eta_{TPV}} \qquad (1)$$

with the full derivation given in Section S3. The first term is the amortized cost of the TPV cells. When the capital cost per area of the cells (CPA, $/cm²) is divided by the power density ($P_{dens,TPV}$, W/cm²), we get the capital cost in $/W-e. We then divide by the time the system outputs electricity per year ($t_{out}$, h) to get the capital cost in $/Wh-e, which can be amortized by the capital recovery factor (CRF). The second term is the cost of providing the heat for the TPV cells, given by the levelized cost of heat (LCOH, $/Wh-th) [26] divided by the TPV cell efficiency ($\eta_{TPV}$). The components of the LCOE equation are summarized in Table 2.

*Table 2: Variables used in calculating the levelized cost of electricity (LCOE) for TPV power generation.*

| Variable | Definition | Units |
|---|---|---|
| LCOE | Levelized cost of electricity | $/Wh-e |
| CPA | TPV cell cost per area | $/cm² |
| CRF | Capital recovery factor | - |
| $P_{dens,TPV}$ | TPV power density | W/cm² |
| CF | Capacity factor | - |
| $t_{out}$ | Operating time over a year | h |
| LCOH | Levelized cost of heating | $/Wh-th |
| $\eta_{TPV}$ | TPV efficiency | - |

With this approach, we can combine the two important TPV performance metrics into one unifying cost metric – LCOE. We can now also directly compare the cost of TPV power generation to other competing options.

While CPA can be taken from literature or manufacturer values, we still need a methodology of calculating LCOH, $P_{dens,TPV}$, and $\eta_{TPV}$. The following sections will discuss how these terms are calculated.

## 2.2. Calculating levelized cost of heating (LCOH)

We calculate LCOH [26] as the cost to provide heat $P_{rad}$ to the TPV cells for time $t_{out}$:

$$LCOH = \frac{CAPEX_{infras}}{P_{rad} t_{out}} \cdot CRF + \frac{OPEX_{input}}{P_{rad} t_{out}}$$

$$\Rightarrow LCOH = \frac{CAPEX_{infras}}{P_{in} \eta_{rad} t_{out}} \cdot CRF + \frac{OPEX_{input}}{P_{in} \eta_{rad} t_{out}}$$

where $CAPEX_{infras}$ is the infrastructure cost of the heating system and $OPEX_{input}$ is the input cost (e.g. fuel or electricity). Note that $\eta_{rad}$ appears in the denominator so a high radiative efficiency is critical to reduce LCOH and therefore LCOE – further discussion is presented in Section S1. We can group terms as:

$$CPE_{th,infras} = \frac{CAPEX_{infras}}{P_{in} \eta_{rad} t_{out}}$$

$$CPE_{th,input} = \frac{OPEX_{input}}{P_{in} \eta_{rad} t_{out}}$$

and therefore:

$$LCOH = CPE_{th,infras} \cdot CRF + CPE_{th,input} \quad (2)$$

where $CPE_{th,infras}$ ($/Wh-th) is the infrastructure cost of the heating system and $CPE_{th,input}$ ($/Wh-th) is the input cost, both per heat energy provided to the TPV cells.

Therefore, if we evaluate the $CPE_{th,infras}$ and $CPE_{th,input}$ for a specific TPV system, then we can use them to calculate LCOH, and plug into Equation 1 to calculate LCOE. Examples of calculating these CPEs are given in Section S2. The components of calculating LCOH are summarized in Table 3.

*Table 3: Variables used in calculating the levelized cost of heating (LCOH) of heat devices for TPV systems.*

| Variable | Definition | Units |
| --- | --- | --- |
| LCOH | Levelized cost of heating | $/Wh-th |
| $CPE_{th,infras}$ | Infrastructure cost for heat device per thermal energy absorbed by the TPV cells | $/Wh-th |
| CRF | Capital recovery factor | - |
| $CPE_{th,input}$ | Input cost per thermal energy absorbed by the TPV cells | $/Wh-th |
| $\eta_{rad}$ | Radiative efficiency (conversion of input energy to thermal energy absorbed by TPV cells) | - |

## 2.3. Calculating TPV cell efficiency and power density

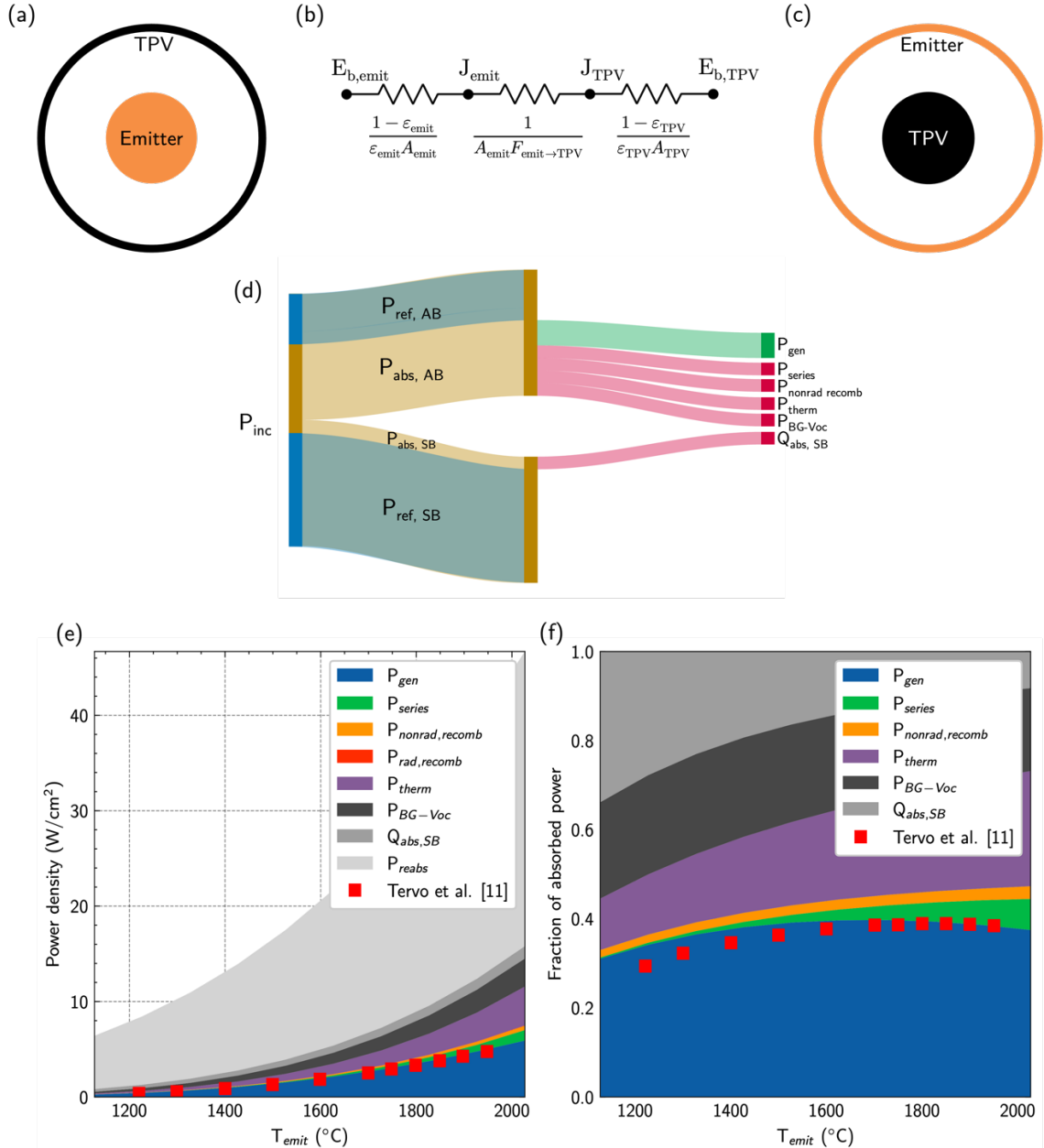

*Figure 2: Components of TPV model and validation results. (a) Emitter-cell configuration used for this work, with low TPV-emitter view factor (b) Radiation network used to calculate net heat flux from emitter to TPV, accounting for emissivity, view factor, and multiple reflections between the surfaces. (c) Emitter-cell configuration suggested for real systems to ensure high TPV-emitter view factor. (d) Sankey diagram showing possible flow paths of incident power $P_{inc}$. $P_{inc}$ includes both above-bandgap (AB) and sub-bandgap (SB) emission regions, and power in each region can either be absorbed (abs) or reflected (ref). Absorbed power above the bandgap generates electron-hole pairs, and loss mechanisms include series resistance $P_{series}$, nonradiative recombination $P_{nonrad,recomb}$, thermalization losses $P_{therm}$, bandgap-open circuit voltage offset $P_{BG-Voc}$. Remaining power is extracted as electricity $P_{gen}$. Absorbed power below the bandgap generates parasitic heating $Q_{abs,SB}$. Also showing validation of the TPV model against experimental data from Tervo et al. [11] for an emitter of various temperatures ($T_{emit}$) with a view factor of 0.31 and an emissivity of 0.9: (e) power density, quantifying values of the different avenues the power can take, and showing the generated power in blue, and (f) efficiency, demonstrating the different loss mechanisms of absorbed power, where the blue section is generated electrical power as a fraction of total power absorbed.*

We develop a model of the TPV cell that determines the efficiency and power density of the cell based on cell properties (bandgap, reflectance, series resistance, and quantum efficiency), and emitter properties (emitter temperature, view factor, emissivity, and bandwidth). We consider a system with TPV cells surrounding an emitter (as shown in Figure 2(a)) to best map to experimental TPV measurements which have low view factor from TPV to emitter – real devices could be inverted with the emitter surrounding the TPV cells to ensure a high view factor (as shown in Figure 2(c)) if allowed by the application. We do not explicitly include a spectral filter, as incorporating a back-surface reflector on the TPV or tuning the emitter emissivity can achieve filtering.

We are interested in power and efficiency so we take a power-based approach to modeling the TPV system similar to Zenker et al [12]. We first start with a radiative network as shown in Figure 2(b) to determine how much power is incident on the TPV cell based on emitter and cell emissivities in the above- and below-bandgap regions. We then calculate whether the incident power is absorbed by the cell or reflected. A Sankey diagram showing possibilities of incident power is shown in Figure 2(d). Absorbed photons above the bandgap generate electron hole pairs, and we include several loss mechanisms (radiative recombination, non-radiative recombination, thermalization, bandgap-Voc offset, and series resistance), with the remaining power being extracted as electricity. Absorbed photons below the bandgap generate parasitic heating and reduce the cell efficiency. Full details of the model are presented in Section S4.

The key assumptions in the model include: (1) cells are single-junction, (2) cell properties are independent of bandgap, (3) cells are opaque, (4) all radiative recombination returns to the emitter, (5) cooling and balance-of-plant power consumption are negligible, and (6) cell cost is independent of cell properties. The implications of these assumptions are discussed in detail at the end of Section 3.3.

We validate the model against the experimental data provided in Tervo et al [11], who were able to demonstrate high efficiency (38.8%) and power density (3.78 W/cm$^2$) using a single-junction cell with a high below-bandgap reflectance (94.7%), low series resistance (6.5 mΩ cm$^2$), and high emitter temperature (1850°C) / high bandgap (0.75eV) to ensure low bandgap-Voc offset (0.2eV). The effects of all these parameters are discussed in detail in Section 3.3. All input parameters for cell and emitter parameters are provided in Table 5, with model validation shown in Figure 2(e) and (f).

With these tools, we can now determine which performance metric (efficiency or power density) is most critical to reducing LCOE for certain applications, and then conduct a sensitivity analysis to determine which cell property change results in the greatest improvement in LCOE.

## 3. Results and Discussion

### 3.1. *General identification of limiting performance metric*

To understand generally how TPV efficiency vs. power density impact the LCOE, we can consider different cost values. As a nominal case, we can use an interest rate of

4% and a lifetime of 20 years such that CRF = 0.074, and let $t_{out} = 8760h$. Then Equation 1 simplifies to

$$LCOE = \frac{CPA \cdot C_1}{P_{dens,TPV}} + \frac{LCOH}{\eta_{TPV}} \tag{3}$$

where $C_1 = \frac{CRF}{t_{out}} = 8.45e{-}6$ 1/h. Then, for specific values of CPA and LCOH, we can determine how power density or TPV cell efficiency impact the LCOE. This is shown in Figure 3(a). When the LCOH is high and TPV cell CPA is low (top left), the heating term (second term) in Equation 3 dominates. High efficiency is critical as it maximizes conversion of heat to electricity, reducing heating costs. This is thus the heating-dominated, efficiency-limited case. In contrast, when the LCOH is low and TPV CPA is high (bottom right), the cell cost term (first term) in Equation 3 dominates. Higher TPV power density Is important, as it reduces the cell area needed for conversion, lowering cell costs. This is the cell-dominated, power-limited case. In the middle, where both terms are equally weighted, both efficiency and power density lower the LCOE, so this is the balanced costs, dual-limited case.

These cases are summarized in the regime map shown in Figure 3(b). As seen, when heating dominates the cost, efficiency is more important, while when cells dominate the cost, power density is more important. Examples of systems in each zone of the regime map are shown in Figure 3(c-d), in each case showing how doubling of the important performance metric can nearly half the LCOE. Regime maps for alternative values of $C_1$ are presented in Section S5, and show different relative importances between efficiency and power density (e.g. as $C_1$ increases, reducing cell cost e.g. improving power density becomes more important).

An interesting observation is that TPV cell efficiency and TPV cell cost per area appear in different terms of the LCOE equation. This suggests that in cases where LCOH dominates, more expensive cells could be used because the cell cost is low compared to heating cost. We quantify this effect In Section S6, finding that in the heating-dominated case, the CPA can increase drastically for small increases in cell efficiency (e.g. a cell with efficiency 0.33 and CPA $18/cm$^2$ has the same LCOE as a cell of efficiency 0.3 and CPA $5/cm$^2$). In contrast, because CPA and power density appear in the same term in the LCOE, a 10% increase in power density must be balanced with a ~10% increase in CPA to retain LCOE.

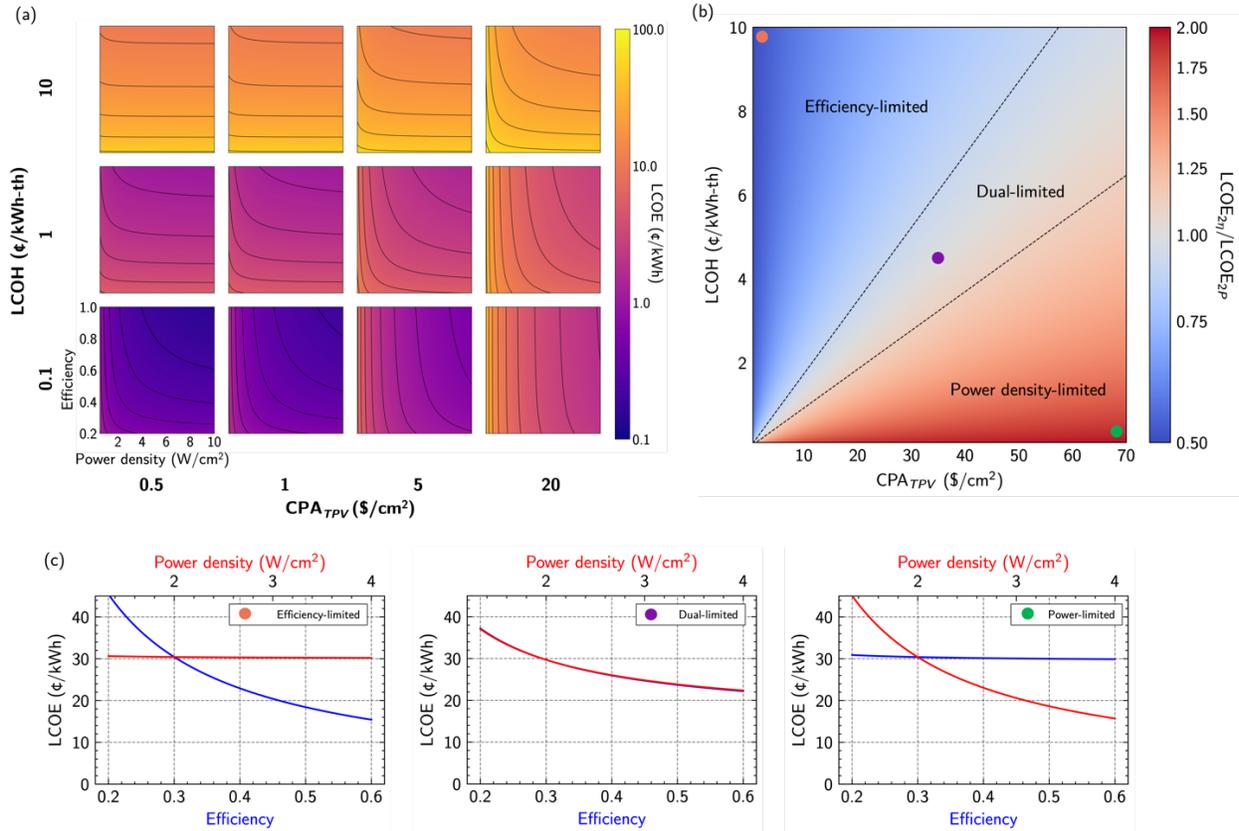

*Figure 3: Regime maps demonstrating the relative importance of efficiency vs. power density in different cost scenarios,* (a) Map of the cost space and how the TPV power density and efficiency impact the LCOE for each set of cost values based on Equation 3. TPV costs per area (CPA) is varied between 0.5, 1, 5, and 20 $/cm² (based on values available in literature) [27,28] and levelized cost per thermal energy (LCOH) is varied between 0.1, 1, and 10 ¢/kWh-th (based on estimates of fuel, electricity, and infrastructure costs) [26]. Columns are constant CPA while rows are constant LCOH. Lowest cost is in the bottom left corner and highest in the top right. Power density values on the x-axis chosen based on the lowest and highest TPV power densities reported in literature [29]. For reference, the LCOE of other electricity generation techniques including natural gas combined cycle (NGCC), advanced nuclear, and advanced coal range from 3 to 10 ¢/kWh-th [30]. (b) Regime map of cost space showing where efficiency or power density is more important to reduce LCOE. From a base case of power density of 2 W/cm² and efficiency of 30%, the LCOE is calculated for a doubled value of each performance metric ($LCOE_{2\eta}$ and $LCOE_{2P}$, as shown in Figure S2(c)). For cases where $LCOE_{2\eta} < LCOE_{2P}$, the efficiency is limiting; when $LCOE_{2P} < LCOE_{2\eta}$, power density is limiting; and when $LCOE_{2\eta} \approx LCOE_{2P}$ both metrics are limiting. (c) Plots of LCOE vs. efficiency (blue line) and LCOE vs. power density (red line) for (left) the effiency limited case where CPA = 1 $/cm² and LCOH = 9 ¢/kWh-th, (center) for the dual limited case where CPA = 35 $/cm² and LCOH = 4.5 ¢/Wh-th, and (right) for the power limited case where CPA = 70 $/cm² and LCOH = 0.3 ¢/kWh-th.

We now know that the importance of TPV power density vs. efficiency in different situations depends on which term in Equation 3 dominates – cell cost or heating cost. However, the cost numbers chosen here may seem arbitrary without concrete examples. Therefore, for the next analysis we consider several specific systems with different applications (and therefore different temperature, capacity factor, and LCOH). Depending on the actual system considered, we can determine which metric is more beneficial to optimize.

## 3.2. Limiting performance metric for example systems

In this work we consider five example systems encompassing common use cases for TPV: (1) solar TPV, (2) waste heat recovery, (3) portable power generation, (4) power plant electricity generation, and (5) thermal storage. Each system is summarized here with full design information provided in Section S2.

For solar TPV, a Fresnel lens is used to concentrate light onto a heat absorber that then radiates light towards a TPV cell [31]. For waste heat recovery, TPV cells are placed around solid hot slag as the output of a cement plant [32,33]. For portable power generation, a microcombustor is used to power TPV cells in a vacuum-sealed package [34]. For power plant electricity generation, a large-scale silicon carbide-based combustor is used to generate heat from hydrogen combustion which then powers TPV cells. For thermal storage, electricity is used to heat graphite blocks to high temperatures, which are then cooled when exposed to TPV cells [3].

For each system, we want to calculate LCOH, which takes $CPE_{th,infras}, CRF, t_{out}$, and $CPE_{th,input}$ as inputs as seen in Equation 2. We calculate $CPE_{th,infras}$ and $CPE_{th,input}$ based on system components, as described in Section S2. We assume all systems have a lifetime of 20 years with a 4% interest rate such that CRF = 0.074 (alternative cases are presented in Section S5). The capacity factor (CF) depends on the application (described in Section S2) and is used to calculate $t_{out} = CF \cdot 8760$ hr. We use these values to calculate LCOH based on Equation 2.

Next, we assume the CPA for the TPV cells for all systems is $5/cm$^2$, typical of high-performing III-V cells [27], but the analysis could be repeated for cells of different types, qualities, or manufacturing methods – in particular, III-V cell cost could be reduced significantly by utilizing substrate reuse, hydride vapor phase epitaxy, and scaling up production [28].

Finally, we can use these values to calculate LCOH (using Equation 2) and the amortized TPV cell cost ($CPA \cdot CRF / t_{out}$), the two cost metrics from Equation 1. These cost metrics for each system are presented in Table 4 (along with all values used as inputs to calculate them) and plotted in Figure 4 to demonstrate their location on the cost regime map presented prior. Note that these are example systems, and their costs are estimates based on information available in literature – system costs can vary significantly based on implementation and these costs are not meant to be prescriptive. However, they can provide a useful survey of the costs of different systems and where they lie in the regime map.

Table 4: Cost values used for the five specific TPV system applications considered in this study. $CPE_{th,system}$ corresponds to all infrastructure costs (capital expenditures) besides the TPV, and $CPE_{th,input}$ corresponds to the input power costs (e.g. fuel, electricity). Capacity factor (CF) of power output also included which is defined as the fraction of total hours in a year the system is outputting electricity.

| Application | $T_{emit}$ (°C) | $CPE_{th,infras}$ (¢/kWh-th) | $CPE_{th,input}$ (¢/kWh-th) | LCOH (¢/kWh-th) | CPA ($/cm$^2$) | CRF | CF | $CPA \cdot \dfrac{CRF}{t_{out}}$ ($/cm$^2$/h) |
|---|---|---|---|---|---|---|---|---|
| Solar TPV | 1400 | 190 | 0 | 14.0 | 5 | 0.074 | 0.20 | 2.11e-04 |
| Waste heat | 1000 | 3.15 | 0 | 0.232 | 5 | 0.074 | 1.00 | 4.22e-05 |

| Portable power | 1200 | 55.1 | 5.86 | 9.91 | 5 | 0.074 | 0.25 | 1.69e-04 |
| Power plant | 1700 | 12.9 | 16.7 | 17.6 | 5 | 0.074 | 1.00 | 4.22e-05 |
| Thermal storage | 2150 | 12.0 | 3.33 | 4.22 | 5 | 0.074 | 0.83 | 5.09e-05 |

Knowing these values, we can now calculate LCOE for different TPV efficiencies or power densities based on Equation 1. However, the full range of efficiencies and power densities may not be achievable by the TPV system (e.g. low temperature of waste heat limits power density). To bound the possible values of these performance metrics, we consider an ideal single-junction TPV cell (100% sub-bandgap reflectance, only thermalization and radiative recombination losses for above-bandgap absorption) with a view factor of 1 to a black emitter. Then, we calculate the bounds for power density and efficiency using the model in Section 2.3, for temperatures representative for the various applications above. The results are presented in **Figure 4**.

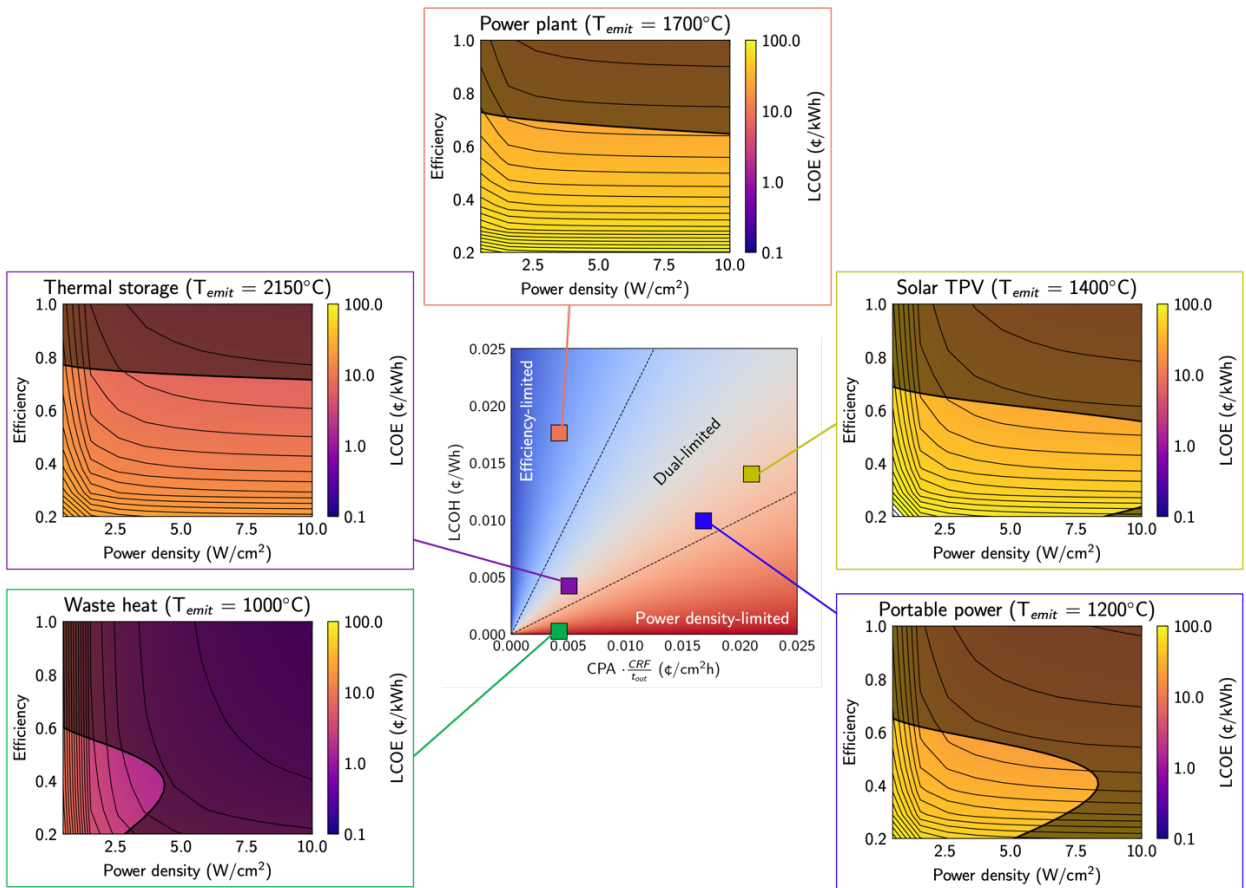

*Figure 4: Where specific TPV applications lie on the efficiency- vs. power-limited regime map.* The cost metrics (heating cost (LCOH) and amortized TPV cell cost ($CPA \cdot CRF / t_{out}$)) for the 5 specific technologies in colored boxes in the central plot. For each technology, also showing how varying the TPV efficiency and power density impact the LCOE using Equation 1. The bounds are determined based on a single-kunction TPV cell with ideal proeprties. All cost metrics given in Table 4. Note that these are example systems, and their costs are estimates based on information available in literature – system costs can vary significantly based on implementation and these costs are not meant to be prescriptive.

As seen, the important performance metric depends on the system. Some applications such as waste heat recovery (green) have little to no energy input costs (waste heat is nearly free) and low infrastructure costs, so the LCOH is low enough that TPV cell costs dominate and power density must be prioritized. Other systems such as the hydrogen-combustion power plant (orange) may use expensive fuels or specialized infrastructure, so LCOH dominates and efficiency must be prioritized. And some systems such as thermal energy storage, solar TPV, and portable power have balanced costs, with cheap electricity as input and cheap infrastructure summing to an LCOH that balances the CPA of the cells, such that both efficiency and power density being important.

Further, we see that the temperature of the application dictates the bounds of TPV efficiency and power density, and therefore LCOE. For example, in waste heat recovery, the low temperature limits the number of photons emitted and therefore power density, and for the power plant, the temperature limits the thermodynamics (e.g. detailed balance) and therefore the maximum efficiency.

As identified in both analyses above, there are 3 cases of interest: efficiency-limited, power-limited, and dual-limited systems, based on whether the cell or heating cost dominates in Equation 3. Clearly in each case we want cells with a high value of the limiting performance metric. While we have presented bounds for performance above, we need to better understand where real TPV cells lie in this region, and which cell property improvements are most important in pushing towards lower LCOE.

### 3.3. TPV improvements to reduce LCOE

We consider 5 cell properties and 5 emitter properties. The cell properties include:
- (i) cell bandgap (BG),
- (ii) back-surface sub-bandgap reflectance (SBR), the reflectance of the back surface of the TPV cell for values below the bandgap energy, assumed to be spectrally uniform,
- (iii) front-surface above-bandgap reflectance (ABR), the reflectance of the top surface of the TPV cell for values above the bandgap energy, assumed to be spectrally uniform,
- (iv) series resistance ($R_{series}$), the resistance to current flow through the TPV active area and metal contacts, and
- (v) non-radiative to radiative recombination ratio (NRR), an empirical parameter comparing the rate of non-radiative recombination processes (e.g. Auger, Shockley-Reed-Hall) to radiative recombination.

The emitter properties include:
- (i) emitter temperature ($T_{emit}$),
- (ii) view factor (VF) from the TPV cell to the emitter,
- (iii) above-bandgap emitter emissivity (ABE), the emissivity of the emitter for values above the bandgap energy,
- (iv) sub-bandgap emitter emissivity (SBE), the emissivity of the emitter for values below the bandgap energy, relevant for spectrally selective emitters, and

(v) emitter bandwidth to cell bandgap ratio (BW), where emitter bandwidth is defined as the range of wavelengths (in nm) below the cell bandgap wavelength that the emitter emits radiation. For example, for a cell bandgap of 1.0 eV (1.24 $\mu$m) an emitter bandwidth from 0 to 1.24 $\mu$m would correspond to a ratio of 1.

The specific cell and emitter properties used for modeling the real cell are presented in Table 5, based on the properties obtained by Tervo et al. [11], along with a comparison to the ideal properties considered earlier.

Table 5: Cell properties used as input for the TPV model for the 2 cells used in this work – real cells as characterized by Tervo et al. [11] and ideal cells with no losses except thermalization. Both are single-junction with selected bandgaps tested in this study (Tervo et al. [11] used a bandgap of 0.75eV).

| Cell property | Abbreviation | Real cell | Ideal cell |
|---|---|---|---|
| Bandgap (eV) | BG | see **Figure 5** | |
| Back-surface (sub-bandgap) reflectance | SBR | 0.947 | 1 |
| Front-surface (above-bandgap) reflectance | ABR | 0.3 | 0 |
| Series resistance ($\Omega$ cm$^2$) | R$_{series}$ | 0.0065 | 0 |
| Nonradiative / radiative recombination ratio | NRR | 12 | 0 |
| **Emitter property** | | | |
| Emitter temperature (°C) | T$_{emit}$ | see Table 4 | |
| View factor | VF | 0.31 | 1 |
| Above-bandgap emitter emissivity | ABE | 0.9 | 1 |
| Sub-bandgap emitter emissivity | SBE | 0.9 | 1 |
| Emitter bandwidth / cell bandgap ratio | BW | 1 | 1 |

With these parameters as input, we use the model described in Sections 2.3 and S4 to calculate cell efficiency vs. power density for various cell bandgaps, where we assume these cell properties are independent of bandgap for simplicity of analysis. We overlay these lines on the cost plots for each of the 3 cases: efficiency-limited, power-limited, and dual-limited, as shown in Figure 5.

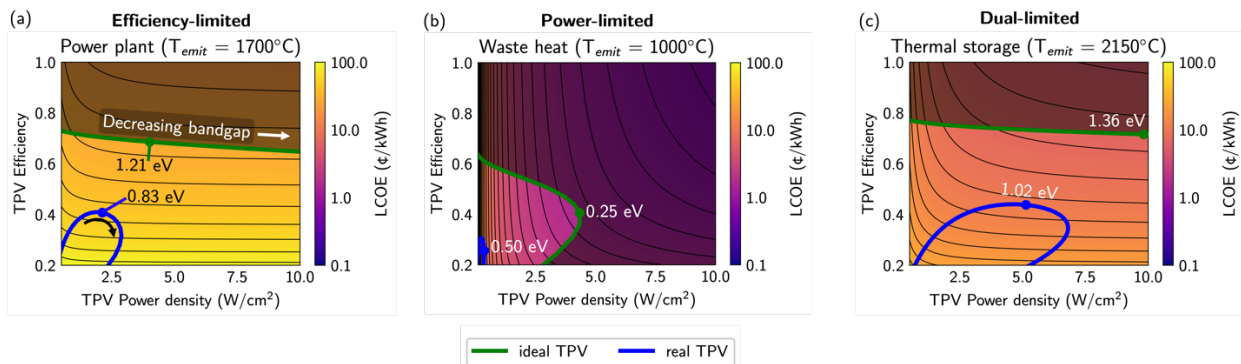

*Figure 5: Colormaps of how TPV power density and efficiency impact the LCOE*, for (a) the efficiency limited case of power-plant scale generation, (b) the power-limited case of waste heat recovery, and (c) the dual-limited case of thermal storage, where cost metrics for LCOH and amortize TPV cell cost are given in Table 4. Also shown for all cases is the efficiency vs. power density curve for an ideal TPV cell (green curve) and a real cell (blue curve) for different values of the bandgap, with the highest bandgap starting on the left and decreasing right. The optimal bandgaps that minimize LCOE for both ideal and real cells are marked with circles. Cell and emitter properties provided in Table 5.

We can see that for both the real and ideal TPVs, the ideal bandgap changes depending on the application. When efficiency-limited, higher bandgaps are favored (e.g. 0.83 eV in **Figure 5**(a)) due to lower thermalization and BG-$V_{oc}$ offset losses leading to higher efficiency. When power limited, lower bandgaps are favored (e.g. 0.50 eV in **Figure 5**(b)) due to greater number of electron-hole pairs generated leading to higher power density. The temperature also impacts the optimal bandgap, due to the emitter's peak wavelength shifting to higher energies at high temperatures per Wien's displacement law, so higher bandgaps are favored (e.g. 1.02 eV in **Figure 5**(c)). We also note that very low bandgaps are often poorly performing, with both worse efficiency and power density. The turnaround at very low bandgaps shown in **Figure 5** is due to either higher recombination (radiative causing larger bandgap-Voc offset relative to bandgap, and non-radiative including Auger or Shockley-Reed-Hall reducing charge carrier collection) or higher series resistance, both resulting in a lower fill factor.

We notice a significant disparity between the performance of a real cell and ideal cell, warranting investigation into cell improvements. The model we developed allows us to test which physical changes to the cell or emitter can most improve the LCOE. We keep the cell bandgap constant at the optimal value for the real cell determined in **Figure 5** and temperature constant at the values in Table 4. Then we vary the 4 remaining cell properties and 4 remaining emitter properties independently to determine their individual effect on LCOE.

On the cell side, we can improve the 4 properties listed previously in Table 5. Increasing the sub-bandgap reflectance means ensuring low-energy photons are recycled and not parasitically absorbed. Reducing front-surface above-bandgap reflectance with e.g. an anti-reflective coating improves the transmission of light into the TPV cell and thus increases electron-hole pair generation. Reducing the series resistance ensures limited heat generation by current flow and is thus particularly important at high current densities. Reducing non-radiative recombination means ensuring more generated electrons are collected by the leads and corresponds to an improved external quantum efficiency.

Figure 6 shows the effect of varying each of these properties on the LCOE achieved for each of the 3 cost cases, with the base values for the real and ideal cells also marked for reference. For this analysis, we have only varied the property of interest and kept all other properties constant at the values described in Table 5. Here we present the LCOE results, while the resulting efficiency and power density for each case are presented in Section S7.

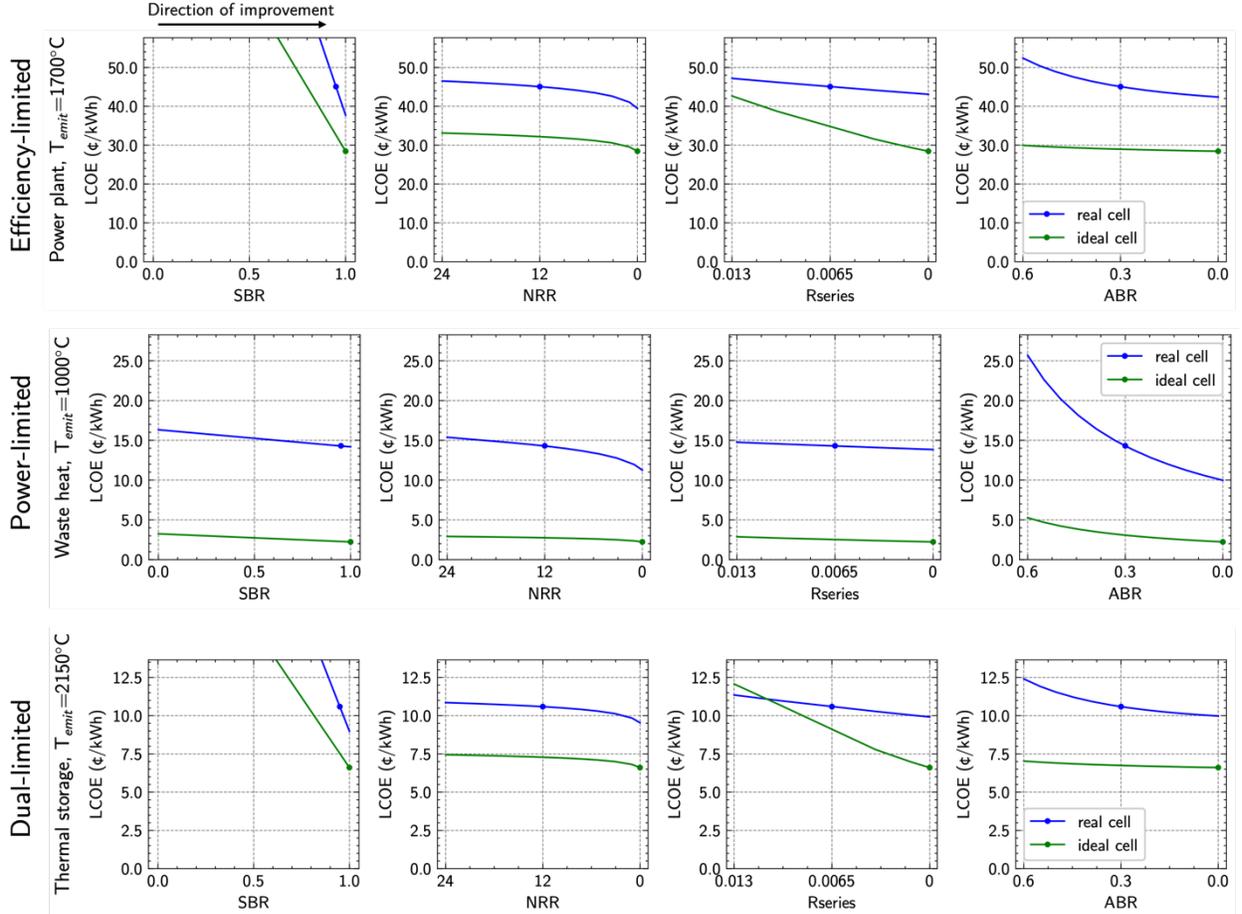

*Figure 6: Effect of varying cell parameters on the LCOE of the system. Columns are back-surface sub-bandgap reflectance (SBR), non-radiative recombination / radiative recombination ratio (NRR), series resistance (Rseries), and front-surface above-bandgap reflectance (ABR). All cell properties besides the one varied are kept constant at the values in Table 5. Top row is the efficiency-limited case, middle row is the power-limited case, and bottom row is the dual-limited case. Blue line shows the real cell while green line shows the ideal cell. Base cases marked with circles. Eefficiency and power density for each case are presented in Section S7, along with the analysis repeated for different emitter temperatures.*

As seen, for the efficiency-limited case, on both the real cell (blue line) and ideal cell (green line) better spectral control by increasing sub-bandgap reflectance (SBR) has a large effect on reducing LCOE. This is because improving the sub-bandgap reflectivity reduces parasitic heating and thus improves efficiency which is the limiting metric for LCOE in this case. Similarly, reducing non-radiative recombination has a large impact since more electron-hole pairs are extracted as electricity instead of being converted to heat, improving efficiency. Series resistance plays a bigger role for the ideal cell which has higher current densities (due to higher view factor and lower above-bandgap reflectance), resulting in larger resistance heating ($I^2 R_{series}$) reducing the efficiency of the cell. Lastly, above-bandgap reflectance primarily improves electrical power generation which has a smaller effect on efficiency, therefore reducing the LCOE to a lesser extent.

For the power-limited case, the above-bandgap reflectance has the greatest impact on LCOE. This is because reducing the above-bandgap reflectance increases the amount of light absorbed by the cell, therefore increasing power density. Non-radiative recombination can also have a similar effect on LCOE, but its effect is muted until very small values of NRR which may be impractical to achieve in real devices. Reducing series resistance has a limited effect in this case due to the lower current densities at low temperatures. Lastly, improving sub-bandgap reflectance has little impact on power output since this primarily reduces parasitic heating and not electricity production.

For the dual-limited case, the same insights from the efficiency-limited case apply, namely importance of SBR and NRR, and the increased impact of series resistance at high view factors and temperatures.

Next, we can change the 4 emitter properties listed previously and in Table 5. Increasing view factor or above-bandgap emissivity uniformly increases the number of photons emitted, increasing the number of electron-hole pairs generated. Reducing the sub-bandgap emissivity improves the spectral selectivity of the emitter and reduces parasitic heating of the TPV cell. Restricting the bandwidth of above-bandgap emission with e.g. a spectral filter also improves spectral selectivity and reduces thermalization losses by suppressing high-energy photon emission.

Figure 7 shows the effect of varying these emitter parameters on the LCOE for the 3 representative cases. Again, all other variables are kept constant at the values described in Table 5, with the resulting efficiency and power density for each case presented in Section S7.

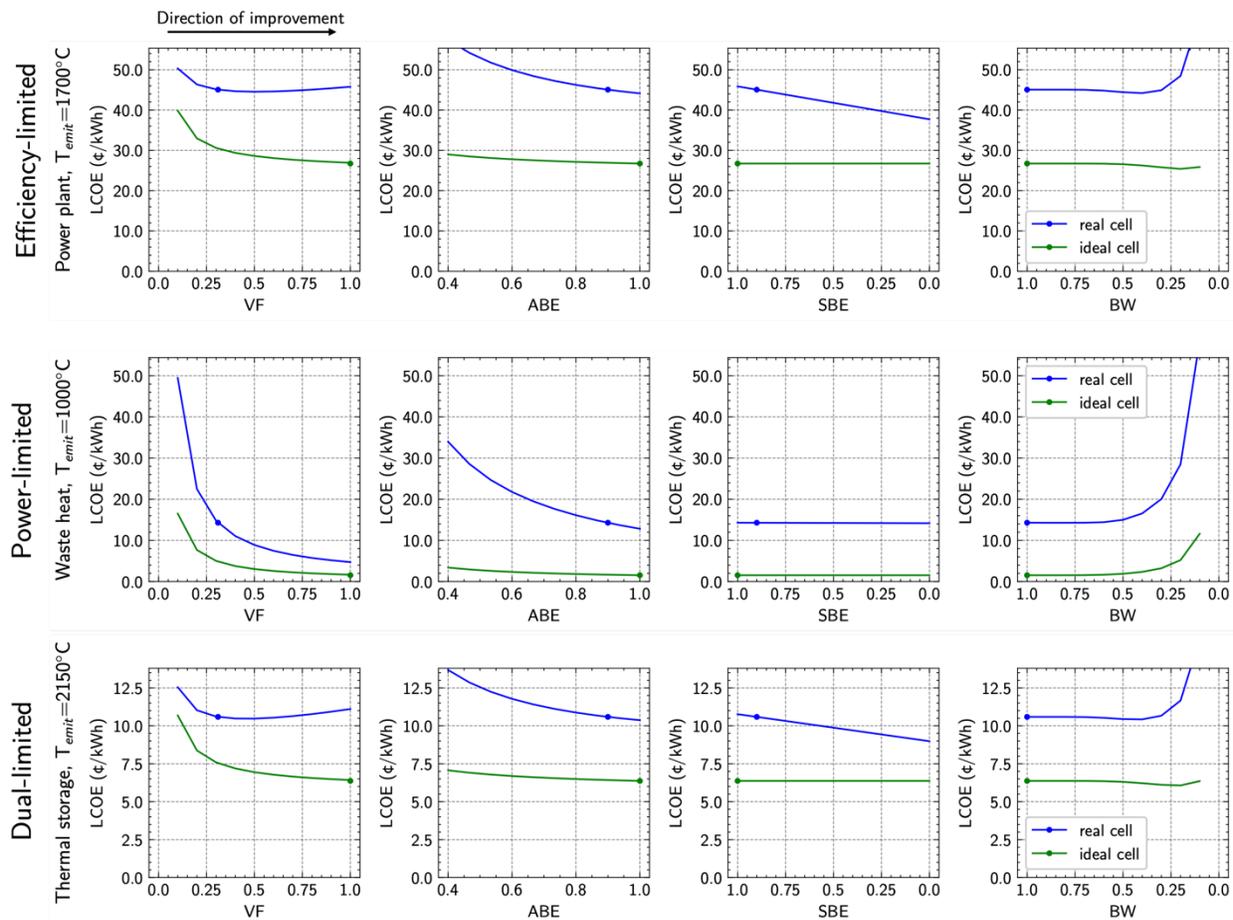

*Figure 7: Effect of varying single emitter parameters on the LCOE of the system.* Columns are TPV cell-to-emitter view factor (VF), above-bandgap emitter emissivity (ABE), sub-bandgap emitter emissivity (SBE), and emitter above-bandgap bandwidth / TPV bandgap ratio (BW). All cell properties besides the one varied are kept constant at the values

*in Table 5. Top row is the efficiency-limited case, middle row is the power-limited case, and bottom row is the dual-limited case. Base case marked with circles. Efficiency and power density for each case are presented in Section S7, along with cases for different emitter temperatures.*

As seen, for efficiency-limited case, reducing the sub-bandgap emitter emissivity (SBE) has the greatest impact on LCOE, but requires reducing emissivity down to 0.1 or less to see benefits comparable to increases in SBR. Further, the value of reducing SBE decreases for higher SBR – for the ideal cell with 100% SBR, changing the SBE has no impact on LCOE. This suggests that improving SBR should be prioritized since smaller changes in SBR (0.95 to 0.99) result in much larger reductions LCOE than large changes in SBE (0.9 to 0.1). Once the SBR has increased to 0.99, which is currently achievable [35], the benefit of reducing SBE is greatly diminished. For other emitter properties, the change is minimal. Increasing view factor or emissivity has negligible impact on the LCOE, and increasing the VF above a certain point can increase the LCOE due to higher series losses in the cell lowering the efficiency.

Interestingly, and perhaps counterintuitively, even in the efficiency-limited case narrowing the above-bandgap bandwidth of the emitter does not significantly change LCOE. Although it does lower the LCOE initially, the LCOE then increases for very narrow bands. This is because of the much lower current density of the cell, which (a) causes a reduction in $V_{oc}$ that outweighs the reduction in thermalization losses, and (b) lowers the power density to a point where TPV costs begin to dominate. We can solve issue (a) by lowering the recombination losses, but even for an ideal cell with only radiative recombination, the impact on LCOE of changing the bandwidth is marginal, as shown in the green line. This suggests both improved quantum efficiency and lower cell cost are required to see the benefits of narrow-bandwidth emission.

For power-limited case, emitter changes can have a large impact. Increasing the view factor can drastically improve power density and therefore LCOE. Improving the view factor of the real cell from 0.31 to 1 results in a similar LCOE to that achieved by an ideal cell with view factor 0.31, indicating a small configurational change can result in a large impact on system performance with no changes to the cell itself. Increasing above-bandgap emissivity also helps, but the starting emissivity is already high. Reducing sub-bandgap emissivity has no impact on power density (only above-bandgap photons create electron-hole pairs), and narrowing the bandwidth only reduces power density, so these changes do not result in LCOE reductions.

For the dual-limited case, again the same insights from the efficiency-limited case apply, namely importance of high above-bandgap emissivity (>0.9) and low sub-bandgap emissivity (<0.1), and limited impact of view factor and bandwidth.

In the above analysis we have identified the effects of single-parameter changes to cell or emitter properties on LCOE, which is useful to identify the most impactful individual changes. In Section S8 we rank the improvements by importance by first setting the highest-impact property to its optimal value, then varying the remaining parameters to identify the next-highest impact property, and repeating this process. The results are consistent with our analysis here, with VF, ABR, and NRR having high importance in the power-limited case and SBR, NRR, and BW having high importance in the efficiency-limited and dual cases. Combining the top two most impactful improvements results in a 27%, 76%, and 24% reduction in LCOE for the efficiency-, power-, and dual-limited cases, respectively.

Our analysis above has assumed a fixed temperature and bandgap for each application. In Section S7, we investigate the impact of temperature on LCOE by repeating the previous analysis (finding the optimal bandgap and then independently varying individual properties) for different emitter temperatures. We find that higher temperature generally results in lower LCOE due to improvements in both efficiency and power density. Efficiency increases because higher bandgaps are optimal, resulting in lower bandgap-$V_{oc}$ offset losses. Power density increases due to greater number of photons hitting the cell at higher temperature since intensity scales with $T^4$. The power-limited case benefits the most from temperature increases, with the LCOE decreasing from 14.3 ¢/kWh at 1000°C to 1.16 ¢/kWh at 2150°C for the base values of the real cell. (For the efficiency-limited case, LCOE decreases from 72.1 to 41.3 ¢/kWh, and for dual-limited from 30.9 to 10.6 ¢/kWh.) We also find that the relative importance of each cell and emitter property is mostly agnostic to different temperatures, however high view factor and emissivity are less important at high temperatures, while low series resistance and low sub-bandgap reflectivity become more important – these altered sensitivities can all be explained by the higher heat fluxes at higher temperatures causing naturally increased power density but also increased parasitic heating.

We can summarize the most effective cell improvements, given a fixed emitter temperature. In efficiency-limited cases, increasing the sub-bandgap reflectance is the most promising option. Reducing the emitter sub-bandgap emissivity helps but requires large changes and high selectivity. Reducing the emitter above-bandgap bandwidth also helps to an extent but must be combined with lower recombination losses and lower cell cost. In power-limited cases, reducing the above-bandgap reflectance and non-radiative recombination while increasing the view factor and emitter emissivity are most impactful. Combining all these improvements helps in dual-limited cases.

The next few paragraphs discuss the primary limitations of our analysis.

In Figures Figure **6** and Figure **7** we have taken all properties to their ideal bounds, but we note that some changes may not be possible or practical. For example, the emitter temperature may be limited by the application (e.g. waste heat has low temperature). Similarly, changing the view factor or adding a spectral filter may not be possible depending on geometric constraints. Lastly, changing the emitter optical properties such as broadband emissivity or selectivity is often material dependent and may not be applicable in all scenarios.

We have assumed the cost metrics (infrastructure cost per energy $CPE_{th,system}$, and input cost per energy $CPE_{th,input}$) are kept constant despite changes in cell or emitter. Infrastructure cost per energy $CPE_{th,system}$ and input cost per energy $CPE_{th,input}$ may change due to second-order effects – for example, a higher power density cell means less cell area is needed for conversion, potentially impacting the emitter surface area as well. Similarly, for storage applications a higher efficiency cell means less thermal energy needs to be stored to generate the same electrical power, impacting infrastructure costs. Another factor could be cooling infrastructure, as systems with lower parasitic heating would require less cooling and therefore could be lower cost. All these effects are application-specific and can be included in more detailed techno-economic models of a specific system.

We assumed certain values of CRF and $t_{out}$ for each system, which influenced the weighting between cell ($CPA \cdot CRF/t_{out}$) and heating (LCOH) costs. For different values of CRF and $t_{out}$, the same system could be in a different cost regime (efficiency vs. power limited) - this is discussed further in Section S5. For example, currently the power plant example system is in the efficiency-limited case, but for sufficiently small $t_{out}$ the cell cost term dominates and the system becomes power-limited, so the resulting impactful cell improvements shift to those for the power-limited case.

As a final caveat for our analysis, we have only considered single-junction cells, but both efficiency and power density can be improved by switching to multi-junction cells. However, multi-junction cells may face similar limitations as the bandwidth analysis, where a lower current density may reduce the $V_{oc}$ for each junction. Reducing recombination losses in multi-junction cells is therefore critical.

## 4. Conclusions and Future Work

In this work we introduced a LCOE-based techno-economic metric to evaluate the relative importance of efficiency and power density in thermophotovoltaic systems. First, we derived the LCOE for TPV systems and found that we can divide the LCOE into two terms – a heating cost term (related to the levelized cost of heating) and cell cost term (related to the amortized cell cost per area). We found that efficiency improvements should be prioritized in systems where the heating cost term dominates, while power density improvements should be prioritized in systems where the TPV cell cost term dominates. In certain cases, the two may be equally important to improve. We then considered five example systems of common TPV applications and noticed that they span 3 limiting cases (efficiency-limited, power-limited, and dual-limited).

This work is thus the first to unify the two important TPV performance metrics – efficiency and power density – through techno-economics in a meaningful way that allows researchers to identify which performance metric is more important for their intended application.

Then, to understand how to maximize the identified important performance metric, we developed a TPV model taking cell and emitter properties as input and predicting the TPV power density and efficiency. Using this model, we conducted a sensitivity analysis by independently varying individual cell variables and noting the improvement in LCOE. For the efficiency-limited case, sub-bandgap reflectance and non-radiative recombination had the highest impact. For the power-limited case, decreasing the above-bandgap reflectance and non-radiative recombination while increasing the view factor helped achieve a low LCOE. Combining the top 2 improvements resulted in a 27% and 76% reduction in LCOE for the efficiency- and power-limited cases, respectively. We also found that increasing the emitter temperature helped for all 3 cases but may not be practical in many applications due to material or input power limitations.

Therefore, the methodology developed in this work enables researchers to identify the most impactful cell or emitter improvements to minimize LCOE, and quantifies the reduction in LCOE expected for maximum enhancement of these properties.

We additionally derived some counterintuitive conclusions of general interest to the TPV community, which we summarize here. First, narrowing the emitter bandwidth in the above-bandgap region had limited impact on LCOE even in the efficiency-limited case – despite the lower thermalization losses, the lower current density reduced the open-circuit voltage. Second, reducing the emitter's sub-bandgap emissivity can only compete with increasing the cell's sub-bandgap reflectivity at very low values of emissivity, which is hard to achieve in practical emitters. Third, in efficiency-limited cases, because the TPV cell cost composes a small part of the overall cost, the cell cost per area can increase significantly to accommodate small improvements in efficiency and still reduce the LCOE.

There are several avenues of future work. Here, we only considered single-junction cells in our modeling, so expanding to multi-junction cells would be a direct extension and could offer insights on how to increase both efficiency and power density by taking advantage of higher-energy photons. Additionally, we use the same CPEs and cell cost per area despite improvements in cell performance, which may not necessarily be the case for at-scale systems, so there is room for future analysis.

We hope that by applying this framework to the wide variety of TPV systems under development, researchers can understand which performance metric is most important for their system, and how they can improve this performance metric with cell improvements.

## Data availability

Data and code are available on GitHub [36].

## Acknowledgements

The authors would like to thank Alina LaPotin, Sean Lubner, and Minok Park for insightful discussions.

## Author contributions

SV: Conceptualization, Formal analysis, Investigation, Methodology, Writing – original draft. KB: Conceptualization, Methodology, Writing – review and editing. AH: Conceptualization, Supervision, Writing – review and editing

## Funding sources

This material is based upon work supported by the National Science Foundation Graduate Research Fellowship under Award No. 2141064.

## Footnotes

Shortly after submission of this work, a preprint on the market viability of TPV systems also based on levelized cost of electricity was published by Datas et al [37]. The authors clarify that the analyses were developed independently.

# Supplementary Information for "Thermophotovoltaic performance metrics and technoeconomics: efficiency vs. power density"

Shomik Verma, Kyle Buznitsky, Asegun Henry

# Nomenclature

*Abbreviations*
| | |
|---|---|
| CPV | Cost per volume ($/m$^3$) |
| HV | Heating value (J/kg) |
| eVF | Effective view factor |
| SB | Sub-bandgap |
| AB | Above-bandgap |
| abs | Absorbed |
| ref | Reflected |

*Variables*
| | |
|---|---|
| $I_0$ | Initial investment ($) |
| $M_t$ | Operations and maintenance expenditures ($) |
| $F_t$ | Fuel expenditures ($) |
| $E_t$ | Electricity generation (Wh) |
| $q_{sun}$ | Incident solar intensity (W/m$^2$) |
| $V$ | Volume (m$^3$) |
| $h_{ch}$ | Charging duration (h) |
| $h_{dis}$ | Discharging duration (h) |
| $F_{TPV \rightarrow emit}$ | View factor from TPV to emitter |
| $F_{emit \rightarrow TPV}$ | View factor from emitter to TPV |
| $A_{emit}$ | Emitter area (m$^2$) |
| $G$ | Irradiance (W/m$^2$) |
| $J$ | Radiosity (W/m$^2$) |
| $\epsilon$ | Emissivity |
| $q$ | Heat flux (W/m$^2$) |
| $R$ | Radiative network resistance (1/m$^2$) |
| $E_b$ | Blackbody irradiance (W/m$^2$) |
| $E$ | Energy (eV) |
| $\eta_{sys}$ | System efficiency |

*Constants*
| | |
|---|---|
| $c$ | Speed of light (m/s) |
| $h$ | Planck's constant (J·s) |
| $k_B$ | Boltzmann constant (eV/K) |
| $e$ | Electron constant |

## S1. Effect of radiative efficiency on LCOH and overall efficiency

We can calculate the overall system efficiency of a TPV system as

$$\eta_{sys} = \frac{P_{elec}}{P_{in}} = \frac{P_{in}\eta_{rad}\eta_{TPV}}{P_{in}} = \eta_{rad}\eta_{TPV}$$

Using the definition of $P_{elec}$ in Figure 1. The radiative efficiency $\eta_{rad}$ is the ratio of (net) radiative power hitting the TPV cells to the input power. Therefore, it includes 3 main loss mechanisms. First is input power to heat conversion, e.g. combustion efficiency where heat can be lost through the exhaust, or solar absorber efficiency where heat can be lost to the environment. Second is heat retention, e.g. heat loss through convection or conduction despite insulation. Third is radiative losses from the emitter, which may have a view factor to the environment.

Figure S1 shows experimentally demonstrated system efficiencies versus TPV efficiencies. As seen, there is a significant disparity between overall system efficiency and TPV efficiency. This is because the radiative efficiency $\eta_{rad}$ has been low in historical devices, while innovations in TPV cells have improved their efficiency.

However, there are easy solutions available for the 3 loss mechanisms that reduce $\eta_{rad}$ discussed above. For input power to heat conversion, device design can reduce heat loss – for example, in combustion systems a recuperator can be used to capture the heat in the exhaust and preheat the inlet reactant streams. Similarly, an IR window can be used in solar TPV systems to prevent heat loss from the absorber. For heat retention, moving to larger scales reduces the surface area to volume ratio and makes insulation easier. For emitter radiative losses, ensuring a high view factor from emitter to TPV reduces heat losses. For example, in the main text we consider a 2D emitter-TPV configuration with the TPV cells surrounding the emitter such that the view factor is 1. In a real-life 3D system of concentric cylinders, the emitter would have some view factor to the environment. This can be solved by capping one end and making the cylinders long, as shown in Figure S1.

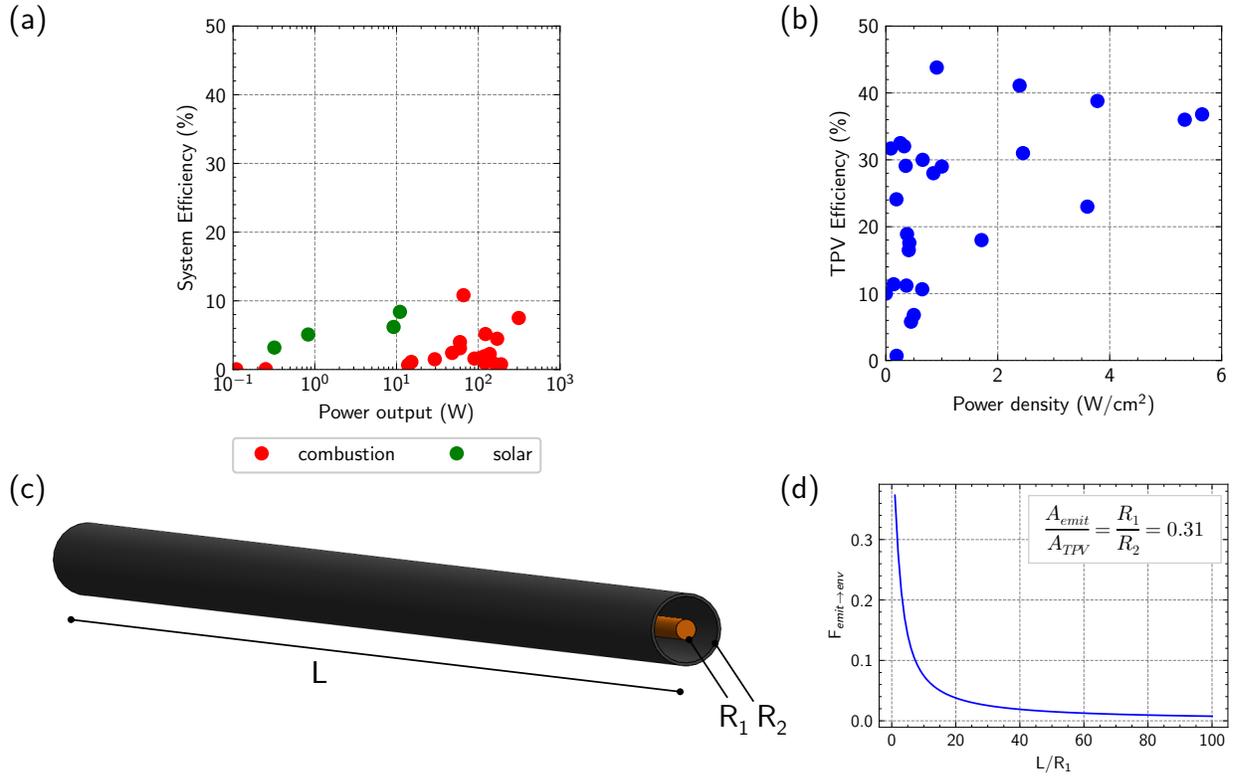

*Figure S1: (a) Overall system efficiencies of combustion and solar-powered TPV systems vs. total output power [24,25]. (b) TPV cell efficiency vs. power density for various cells tested individually [26]. (c,d) Demonstration of reducing the radiative heat loss to the environment component of $\eta_{rad}$. (c) Schematic of emitter – TPV system with cylindrical emitter of radius $R_1$ surrounded by TPV cells with inner radius $R_2$. Area ratio of emitter to TPV is 0.31. (d) View factor from emitter to environment for various ratios of $L/R_1$.*

With these improvements, we can improve $\eta_{rad}$, which is important since LCOH is inversely proportional to $\eta_{rad}$:

$$LCOH = \frac{1}{\eta_{rad}}\left(\frac{CAPEX_{infras}}{P_{in}t_{out}} \cdot CRF + CPE_{input}\right)$$

If the LCOH is already low compared to the TPV cell cost, then improving $\eta_{rad}$ has limited impact on LCOE. However, in many cases LCOH is high, so high $\eta_{rad}$ allows achieving the lower LCOH required for cost-competitive TPV systems. Note that $CAPEX_{infras}$ may increase with higher $\eta_{rad}$ (e.g. with more insulation or better quality materials) so a full trade-off analysis should be done for specific systems.

## S2. Technoeconomic details of specific systems considered

In this section we calculate the $CPE_{th,infras}$ and $CPE_{th,input}$ for 5 example systems. Recall that $CPE_{th,infras}$ [$/Wh-th] is the infrastructure cost of the heat device and $CPE_{th,input}$ ($/Wh-th) is the input cost, both per thermal energy provided to the TPV cells:

$$CPE_{th,infras} = \frac{CAPEX_{infras}}{P_{rad}t_{out}}$$

$$CPE_{th,input} = \frac{OPEX_{input}}{P_{rad}t_{out}}$$

Where $CAPEX_{infras}$ is the total infrastructure capital cost, $OPEX_{input}$ is the operational expense of input power, $P_{rad}$ is the (net) radiative power on the cell, and $t_{out}$ is the number of hours per year the heating system provides power to the TPV cells. In most applications, $P_{rad} = P_{in}\eta_{rad}$ where $P_{in}$ is the input power and $\eta_{rad}$ is the radiative efficiency.

For the solar TPV application we use the system presented in Bhatt and Gupta of a Fresnel lens solar concentrator concentrating light on a shielded solar absorber, using a selective emitter and spectral filter for radiative transfer to the TPV cell [2]. As a solar device we use a capacity factor of 0.2. Because sunlight is used as input, $CPE_{th,input} = 0$. $CPE_{th,infras}$ is again the cost of the system (besides TPV) divided by the thermal energy output which is

$$CPE_{th,infras} = \frac{CAPEX_{infras}}{P_{in}\eta_{rad}t_{out}} = \frac{Cost(device) + Cost(insulation)}{P_{in}\eta_{rad}t_{out}}$$
$$= \frac{Cost(Lens) + Cost(shield) + Cost(absorber) + CPV_{insulation}V_{insulation}}{q_{sun}\eta_{rad}t_{out}}$$

We obtain cost estimates for the lens ($100), shield ($10), absorber ($15), and insulation ($1000/m³) from the literature [3–6]. We use a $\eta_{rad}$ of 75%, $q_{sun}$ of 100W, and $V_{insulation}$ of 0.125m³ for this sized system. This results in a $CPE_{th,infras}$ of 1.90e2 ¢/kWh-th. Larger systems may have lower insulation costs due to a lower surface area to volume ratio.

For waste heat recovery we use the waste heat from hot solid product streams such as cement slags or hot steels, which are often at high temperature and predominantly lose heat through radiation [7,8]. We install a TPV array surrounding the hot solid product stream. As many industrial processes operate near-continuously we use a capacity factor of 1 and we again assume $CPE_{th,input} = 0$. $CPE_{th,system}$ is calculated as

$$CPE_{th,infras} = \frac{CPV_{piping}V_{piping} + CPV_{insulation}V_{insulation}}{P_{in}\eta_{rad}t_{out}}$$

We obtain cost estimates for the piping ($16000/m³), and insulation ($2000/m³) from the literature [9–11]. We use a $\eta_{rad}$ of 90%, $P_{in}$ of 1MW, $V_{piping}$ of 10m³ and $V_{insulation}$ of 50m³ for this large-scale system. This results in a $CPE_{th,infras}$ of 3.16e0 ¢/kWh-th.

For portable power generation we use a system similar to the one presented by Chan et al. using propane as the fuel to power an Inconel microcombustor in a vacuum-sealed package [12]. The capacity factor can vary depending on the application, for this analysis we used $CF = 0.25$ but this could be lower, weighting the infrastructure costs higher than the operating costs. We can then calculate the two CPEs:

$$CPE_{th,input} = \frac{OPEX_{input}}{P_{in}\eta_{rad}t_{out}} = \frac{Cost\ per\ kg_{fuel} \cdot kg_{fuel}}{kg_{fuel}HV_{fuel}\eta_{rad}} = \frac{Cost\ per\ kg_{fuel}}{HV_{fuel}\eta_{rad}}$$

$$CPE_{th,infras} = \frac{Cost(device) + Cost(insulation)}{P_{in}\eta_{rad}t_{out}}$$
$$= \frac{CPV_{device}V_{device} + CPV_{insulation}V_{insulation}}{\dot{m}_{fuel}HV_{fuel}\eta_{rad}t_{out}}$$

We obtain cost estimates for the propane ($1/gal), Inconel ($350000/m3), and insulation ($100/m3) from the literature [13–15]. We use an $\eta_{rad}$ of 70% given higher potential heat losses at small scales, $P_{in}$ of 100W, $V_{device}$ of 2.5e-4 m3, and $V_{insulation}$ of 10e-4 m3. This results in a $CPE_{th,input}$ of 5.86e0 ¢/kWh-th and a $CPE_{th,infras}$ of 5.51e+1 ¢/kWh-th.

Next, we consider power plant-scale power generation with a modular hydrogen combustion device made of silicon carbide, with a 12" combustion zone and a 12" recuperation zone. This enables high combustion efficiency and also scalability where thousands of these could be placed adjacently to reach the MW scale. We can calculate the two CPEs using the same formulas as the portable power generator.

We obtain cost estimates for the hydrogen ($5/kg), silicon carbide ($500000/m3), and insulation ($2000/m3) from the literature [10,11,16]. We use an $\eta_{rad}$ of 90%, $P_{in}$ of 250MW, $V_{device}$ of 500 m3, and $V_{insulation}$ of 1500 m3. This results in a $CPE_{th,input}$ of 1.67e1 ¢/kWh-th and a $CPE_{th,infras}$ of 1.29e1 ¢/kWh-th.

For the thermal storage application we use the thermal energy grid storage system as presented in Kelsall et al. using graphite as the storage medium and liquid tin as the heat transfer fluid [17]. Liquid tin is heated by Joule heating when electricity is cheap, used to charge the graphite blocks, and the graphite blocks release heat towards the TPV when needed. We use the technoeconomics previously presented for the 1GWh-th system with 4 hours of charging ($h_{in}$) and 20 hours of discharging ($h_{out}$).

In this case, $P_{rad}$ cannot be calculated directly as $P_{in}\eta_{rad}$ since the input and output heat transfer are different due to the different durations of charge/discharge. In this case, from conservation of energy we have $P_{in}h_{in}\eta_{rad} = P_{rad}h_{out} \Rightarrow P_{rad} = P_{in}\left(\frac{h_{in}}{h_{out}}\right)\eta_{rad}$.

We can now calculate the two CPEs as:

$$CPE_{th,input} = \frac{Cost\ per\ Wh_{elec}}{\eta_{rad}}$$

$$CPE_{th,infras} = \frac{Cost(storage) + Cost(charging) + Cost(discharge - TPV)}{P_{in}\left(\frac{h_{ch}}{h_{dis}}\right)\eta_{rad}t_{out}}$$

$$= \frac{CPE_{th}(1GWh) + CPP_{ch}(250MW) + CPP_{dis}(45MW\eta_{TPV})}{P_{in}\left(\frac{4}{20}\right)\eta_{rad}t_{out}}$$

We obtain cost estimates for the $CPE_{th}$ ($20/kWh-th), $CPP_{ch}$ ($0.03/W), $CPP_{dis}$ excluding TPV ($0.42/W), and electricity cost ($0.03/kWh) from the previous study, which goes into full detail about how all components of the system contribute to the cost metrics. We use an $\eta_{rad}$ of 0.9, a $P_{in}$ of 250MW-e, and based on the rated durations $P_{out}$ is 45 MW-th. This results in a $CPE_{th,input}$ of 3.33e0 ¢/kWh-th and $CPE_{th,infras}$ of 1.20e1 ¢/kWh-th.

Note that the cost values here are determined for a specific sized system with a certain $P_{in}$. Depending on the cell characteristics, this $P_{in}$ can change. For example, consider a constant emitter temperature $T_{emit}$. Then the power incident on the cell is $P_{inc} = \varepsilon\sigma T_{emit}^4$ (assuming an effective view factor of 1). Consider cell 1 that reflects most of this light such that $P_{ref,1} = 0.9\varepsilon\sigma T_{emit}^4$, then $P_{in,1}\eta_{rad} = P_{inc} - P_{ref,1} = 0.1\varepsilon\sigma T_{emit}^4$. Next consider cell 2 that absorbs most of this light such that $P_{ref,2} = 0.2\varepsilon\sigma T_{emit}^4$ and $P_{in,2}\eta_{rad} = 0.8\varepsilon\sigma T_{emit}^4$. Because cell 2 absorbs more light, the input power is greater. These two systems thus have different $P_{in}$ values due to the different properties of the cell. In this work we have assumed that $CPE_{th,infras} = \frac{CAPEX_{infras}}{P_{in}\eta_{rad}t_{out}}$ remains constant despite changes in cell properties, which means that $CAPEX_{infras}$ scales proportionally with $P_{in}$. This may not be the case for all technologies (e.g. for many thermal systems $CPE_{th,infras}$ decreases for larger sizes due to lower insulation costs [17]) so detailed system-specific technoeconomic models should take this variation into consideration when optimizing TPV cell properties.

The five systems above provide concrete examples of how cost values can be determined for use in the LCOE metric. While they are specific to certain systems, the same methodology can be used for any system of interest, demonstrating the versatility and utility of this metric.

## S3. Derivation of TPV LCOE

The full form of the LCOE is [1]:

$$LCOE = \frac{I_0 + \sum_{t=0}^{n} \frac{M_t + F_t}{(1+i)^t}}{\sum_{t=0}^{n} \frac{E_t}{(1+i)^t}}$$

where $I_0$ is the initial investment or capital expenditure of the project, $M_t$ is the operations and maintenance expenditures in year t, $F_t$ is the fuel expenditures in year t, $E_t$ is the electricity generation in year t, and $i$ is the interest rate. Assuming a constant value of maintenance and fuel expenditures and energy generation year-over-year, this expression simplifies to

$$LCOE = \frac{I_0 \cdot CRF}{E_{out}} + \frac{M + F}{E_{out}}$$

Where CRF is the capital recovery factor defined as $\frac{i(1+i)^n}{(1+i)^n - 1}$ where $n$ is the lifetime in years and $i$ is the interest rate. We can adapt this LCOE for our TPV system:

$$LCOE = \frac{(CAPEX_{TPV} + CAPEX_{infras}) \cdot CRF}{P_{elec} t_{out}} + \frac{OPEX}{P_{elec} t_{out}}$$

We call the input power $P_{in}$ and the input-to-radiative heat conversion efficiency as $\eta_{rad}$, as in Figure 1. Then the output heat power is $P_{in}\eta_{rad}$ and the heat is converted to electricity at efficiency $\eta_{TPV}$ resulting in an output electrical power $P_{elec} = P_{in}\eta_{rad}\eta_{TPV}$. Then, dividing $P_{elec}$ by the power density of the cells determines the area of TPV required, which we can multiply by the cost per area to get the TPV CAPEX, $CAPEX_{TPV} = \frac{P_{elec}}{P_{dens}} CPA$ (where CPA is the cost per area of the TPV cells). We then have

$$LCOE = \frac{CPA \cdot CRF}{P_{dens,TPV} t_{out}} + \frac{CAPEX_{infras} \cdot CRF}{P_{elec} t_{out}} + \frac{OPEX}{P_{elec} t_{out}}$$

We can plug in the expression of $P_{out}$ to get

$$LCOE = \frac{CPA \cdot CRF}{P_{dens,TPV} t_{out}} + \frac{CAPEX_{infras} \cdot CRF}{P_{in}\eta_{rad}\eta_{TPV} t_{out}} + \frac{OPEX}{P_{in}\eta_{rad}\eta_{TPV} t_{out}}$$

Now we can group cost parameters into a cost per energy: $CPE_{th,infras} = \frac{CAPEX_{infras}}{P_{in}\eta_{rad}t_{out}}$ and $CPE_{th,input} = \frac{OPEX}{P_{in}\eta_{rad}t_{out}}$ which are the costs per thermal energy output of the heat system in $/Wh-th. Plugging these terms into the LCOE equation, we get:

$$LCOE = \frac{CPA \cdot CRF}{P_{dens,TPV} t_{out}} + \frac{CPE_{th,infras} \cdot CRF + CPE_{th,input}}{\eta_{TPV}}$$

which is Equation 1 with the definition of LCOH in Equation 2 substituted in.

Note that for our purposes we have only considered energy input costs for the OPEX such that $\frac{OPEX}{P_{in}t_{out}}$ is the cost per energy input (e.g. $/kJ for fuel, $/kWh for electricity, etc). Maintenance costs could be easily added for future studies by increasing the OPEX value by e.g. 1% of the infrastructure costs.

## S4. Details of thermophotovoltaic model

As we are interested in power and efficiency we take a power-based approach to modeling the TPV system similar to Zenker et al [18]. To best map to existing TPV devices measurements, we consider an emitter-cell configuration where the TPV cells surround the emitter, as shown in Figure 2(a). The view factor traditionally reported in TPV measurements (VF) is the view factor from TPV to emitter ($F_{TPV \to emit}$) [19,20], such that

$$A_{TPV} F_{TPV \to emit} = A_{emit} F_{emit \to TPV}$$

$$\Rightarrow VF \approx \frac{A_{emit}}{A_{TPV}}$$

where we have assumed $F_{emit \to TPV} \approx 1$. Further discussion of $F_{emit \to TPV}$ and its impact on radiative efficiency is presented in Section S1.

Then we can create a radiation network as shown in Figure 2(b) to calculate the relevant radiation fluxes. In particular, we are interested in the irradiance $G_2$ (incident radiant heat flux on the TPV surface). From the radiation network we can calculate the net heat flux $q_{emit \to TPV}$ as

$$q_{emit \to TPV} = \frac{E_{b,emit} - E_{b,TPV}}{R_{emit} + R_J + R_{TPV}}$$

Where $R_{emit} = \frac{1-\varepsilon_{emit}}{\varepsilon_{emit} A_{emit}}$, $R_J = \frac{1}{A_{emit} F_{emit \to TPV}}$, and $R_{TPV} = \frac{1-\varepsilon_{TPV}}{\varepsilon_{TPV} A_{TPV}}$. Since $E_{b,emit} \gg E_{b,TPV}$ we can let $E_{b,TPV} = 0$. Once we know $q_{emit \to TPV}$ we can similarly calculate radiosities $J_{emit}$ and $J_{TPV}$ from the network. Finally, we can calculate TPV irradiance $G_{TPV}$ from $-q_{emit \to TPV} = A_{TPV}(J_{TPV} - G_{TPV})$ giving

$$G_{TPV} = \frac{\frac{E_{b,emit}}{\varepsilon_{TPV} A_{TPV}}}{R_{emit} + R_J + R_{TPV}} = E_{b,emit} \cdot eVF$$

We have extracted the effective view factor $eVF = \frac{1}{\varepsilon_{TPV} A_{TPV}(R_{emit} + R_J + R_{TPV})}$ which when multiplied by the blackbody radiation emitted from the emitter gives the incident radiant heat flux on the TPV surface, taking into account both emission from the emitter and reflected light from adjacent TPV cells.

Due to the differing optical properties of the cell below and above its bandgap, we can split this radiation power and calculate the power incident on the cell in the sub- and above-bandgap regions. We split $E_{b,emit}$ using Planck's law and create distinct radiation networks for the two bands to define separate effective view factors for the sub-bandgap ($eVF_{SB}$) and above-bandgap ($eVF_{AB}$) regions.

$$P_{SB} = \frac{2\pi}{c^2 h^3} \int_0^{E_g} \frac{E^3}{e^{\left[\frac{E}{k_B T_{rad}}\right]} - 1} dE \cdot eVF_{SB}$$

$$P_{AB} = \frac{2\pi}{c^2 h^3} \int_{E_g}^{E_{max}} \frac{E^3}{e^{\left[\frac{E}{k_B T_{rad}}\right]} - 1} dE \cdot eVF_{AB}$$

where $E_{max} = \frac{E_g}{1-BW}$. Then, we can use the optical properties of the cell (reflectance) to calculate the reflected vs. absorbed light in each wavelength region.

$$P_{SB,ref} = P_{SB} \cdot SBR$$
$$P_{SB,abs} = P_{SB} - P_{SB,ref}$$
$$P_{AB,ref} = P_{AB} \cdot ABR$$
$$P_{AB,abs} = P_{AB} - P_{AB,ref}$$

Sub-bandgap absorption by the cell cannot generate electron-hole pairs so is considered parasitic heating. Above-bandgap absorption generates electron-hole pairs, but the number or potential of these pairs may reduce before they can be extracted. We consider various loss mechanisms. First is radiative recombination, which is thermodynamic and based on thermal equilibrium of the cell – it can be calculated based on the cell's operating voltage as:

$$P_{rad-recomb} = \frac{2\pi}{c^2 h^3} eV \int_{E_g}^{E_{max}} \frac{E^2}{e^{\left[\frac{E-eV}{k_B T_{cell}}\right]} - 1} dE \cdot eVF_{AB} \cdot (1 - ABR)$$

The next loss mechanism is non-radiative recombination, which can occur from a variety of processes including Shockley–Read–Hall recombination, Auger recombination, or surface recombination. Instead of modeling these processes explicitly, we use an empirical parameter NRR (ratio of non-radiative recombination rate to radiative recombination rate) to calculate the non-radiative recombination.

$$P_{nonrad-recomb} = P_{rad-recomb} \cdot NRR$$

While the recombination terms above (primarily) impact the current density by reducing the number of electrons extracted, two additional terms impact the energy carried by those electrons. Thermalization of charge carriers occurs as they relax to the cell's bandgap. Charge carriers lose further energy because the open-circuit voltage is lower than the bandgap due to recombination [21]. These two terms are calculated as:

$$P_{therm} = \frac{2\pi}{c^2 h^3} \int_{E_g}^{E_{max}} \frac{E^2(E - E_g)}{e^{\left[\frac{E}{k_B T_{rad}}\right]} - 1} dE \cdot eVF_{AB} \cdot (1 - ABR)$$

$$P_{BG-V_{oc}} = \frac{2\pi}{c^2 h^3} \int_{E_g}^{E_{max}} \frac{E^2(E_g - eV)}{e^{\left[\frac{E}{k_B T_{rad}}\right]} - 1} dE \cdot eVF_{AB} \cdot (1 - ABR)$$

Finally, we can calculate the power density as the total absorbed power minus these loss mechanisms, and efficiency as the power density over total absorbed power.

$$P_{gen} = \left(P_{AB} - P_{AB,ref} - P_{therm} - P_{BG-V_{oc}} - P_{rad,recomb} - P_{nonrad,recomb}\right) - \left(\frac{P_{gen}}{V}\right)^2 R_{series}$$

$$\eta_{TPV} = \frac{P_{gen}}{P_{AB} - P_{AB,ref} + P_{SB} - P_{SB,ref} - P_{rad,recomb}}$$

where we operate the cell at the maximum power point. Here we have assumed that radiative recombination does not impact the radiative network (e.g. $P_{rad,recomb} \ll P_{AB,ref} + P_{SB,ref}$) but this contribution could be incorporated into the effective view factor $eVF$ if radiative recombination plays a larger role in future devices. Figure 2(e) and (f) shows a validation of this methodology with a breakdown of where the incident power ends up. The cell parameters used are presented in Table 4. A webapp based on this model is available on GitHub and online [22,23].

## S5. Sensitivity of costs to input parameter values

From Equations 1 and 2, we see that both LCOE and LCOH are sensitive to the CRF which depends on the interest rate $i$ and lifetime $n$:

$$CRF = \frac{i(1+i)^n}{(1+i)^n - 1}$$

The sensitivity of CRF to $i$ and $n$ is plotted in Figure S2(a). Previously in this work, we assumed an interest rate of 4% and lifetime of 20 years, which resulted in a CRF of 0.074. We find that for lifetimes between 10 and 30 years, and interest rates between 1 and 10%, the CRF can vary between 0.03 and 0.14.

This can impact the relative importance of power density vs. efficiency, because only the cell and infrastructure costs are impacted by the CRF (not energy input cost which is operational). Therefore, for higher values of CRF, as shown in Figure S2(b), power density has a higher importance than the base case in Figure 3. Similarly, at lower values of $t_{out}$, power density has higher importance. This is shown explicitly in Figure S2(c), where the base case LCOE is shown (for efficiency 30% and power density 2 W/cm2), then how the LCOE changes if the efficiency vs. power density is doubled.

For LCOH, a higher CRF can similarly weigh the system cost higher than the input energy cost. As shown in Figure S2(d), higher values of CRF (e.g. 0.15 instead of 0.074) mean the initial capital cost of the system should be lower to ensure low LCOH.

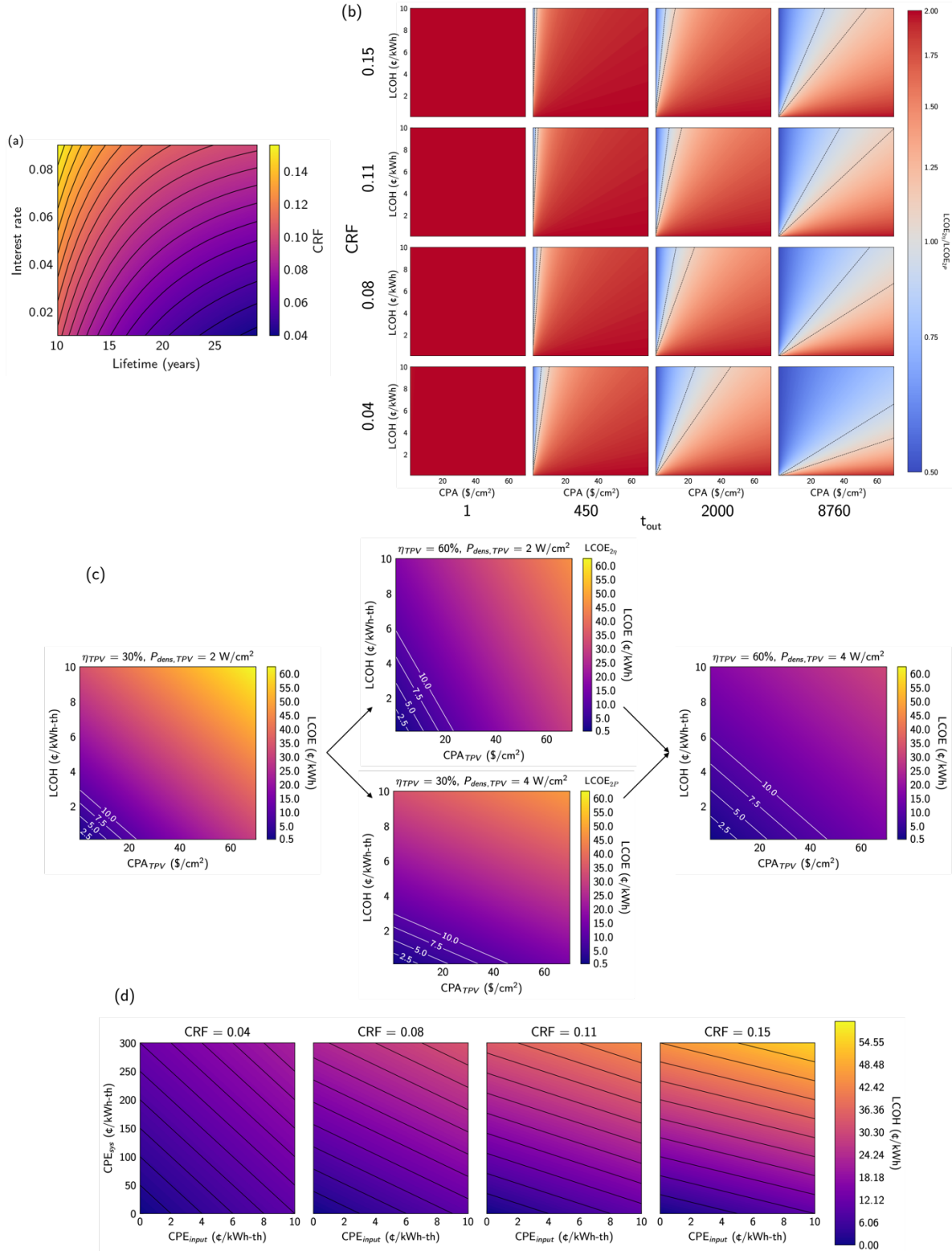

*Figure S2: Sensitivity analysis of costs to input parameters. (a) Sensitivity of CRF to interest rate and lifetime. (b) Sensitivity of efficiency vs. power-limited regime map to different values of CRF and tout. Blue areas are efficiency-limited while red is power-limited. (c) Base-case LCOE, LCOE with double efficiency vs. double power density, and LCOE with double of both metrics. (d) Sensitivity of LCOH computed from Equation 2 to CRF value used.*

## S6. Allowable TPV cell cost per area increase for performance improvements

It is likely that higher performing cells also have a higher cost per area. Therefore, it is important to understand, for a given improvement in cell performance, how much increase in cost per area is acceptable by the TPV system. We again consider our 3 cases of efficiency-limited, power-limited, and dual-limited systems and consider a base TPV cell with efficiency 0.3, power density 2 W/cm$^2$, and CPA \$5/cm$^2$. Then, for a given percent increase in cell performance, we calculate the percentage the CPA can increase to keep the LCOE constant. Any cell improvement that results in a lower CPA than the one calculated would result in a lower overall LCOE. The results of this analysis are shown in Figure S3.

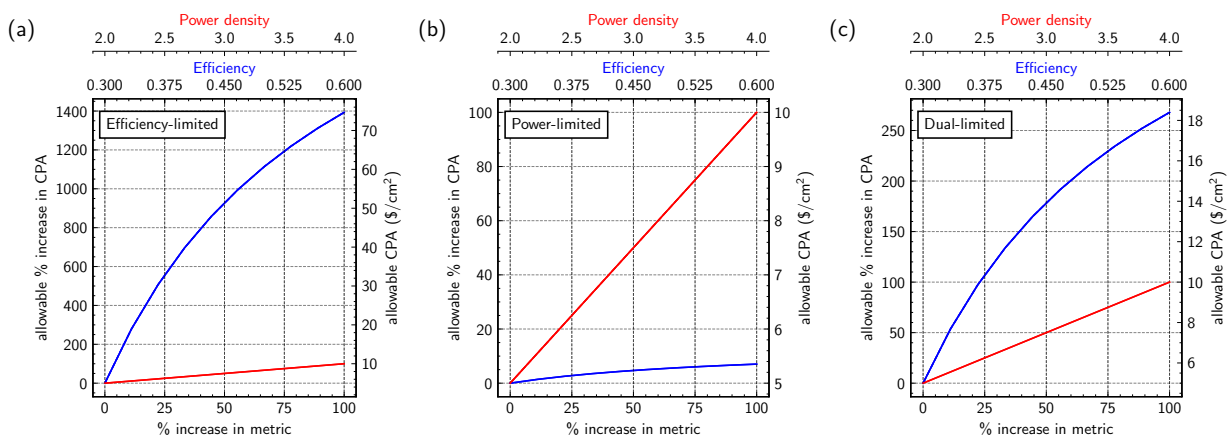

*Figure S3: Allowable increase in CPA to preserve LCOE for % increase in performance metric (efficiency or power density) for three different cases: (a) efficiency-limited, (b) power-limited, and (c) dual-limited, where cost values are the same as in Figure 3(c-f). Efficiency lines given in blue and power density lines in red. For efficiency improvements, power density is kept at the base value of 2 W/cm2, and for power improvements, efficiency is kept at the base value of 0.3.*

First, looking at the power density lines in red, a cell with x% higher power density can be at most x% costlier per area to preserve the LCOE. This is evident from Equation 1 where cost per area is divided by power density, so they must scale equally.

More interesting are the efficiency lines in blue. For the efficiency limited case, we see that there is a large margin for cell costs for small increases in efficiency. For example, a cell with 10% higher efficiency (from 0.3 to 0.33) can be approximately 260% more expensive (\$18/cm$^2$ from \$5/cm$^2$) to preserve the LCOE. This is because the TPV cost is small in comparison to the rest of the system, so higher CPA does not significantly impact the cost. However, higher efficiency drastically reduces the cost, making the benefit outweigh the cost. In contrast, for power-limited cases, even a cell with 100% higher efficiency (0.6) must have a low CPA (\$5.4/cm$^2$) to preserve LCOE.

## S7. Effect of changing cell and emitter variables on efficiency and power density for various emitter temperatures

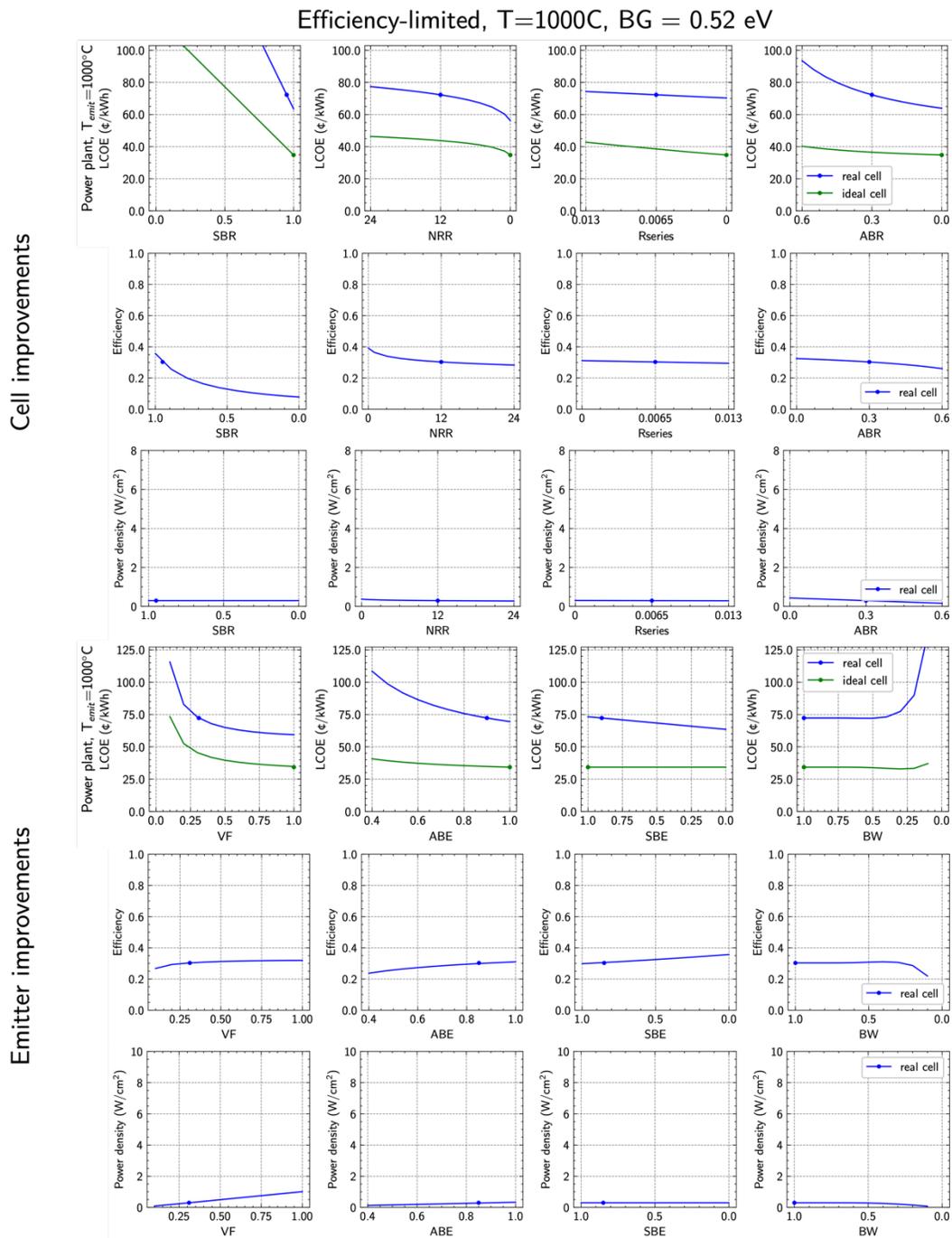

*Figure S4: LCOE, efficiency and power density for the efficiency-limited case with emitter temperature 1000C and cell bandgap 0.52eV. Model of a real cell with properties provided in Table 4 while varying a single property on each x-axis.*

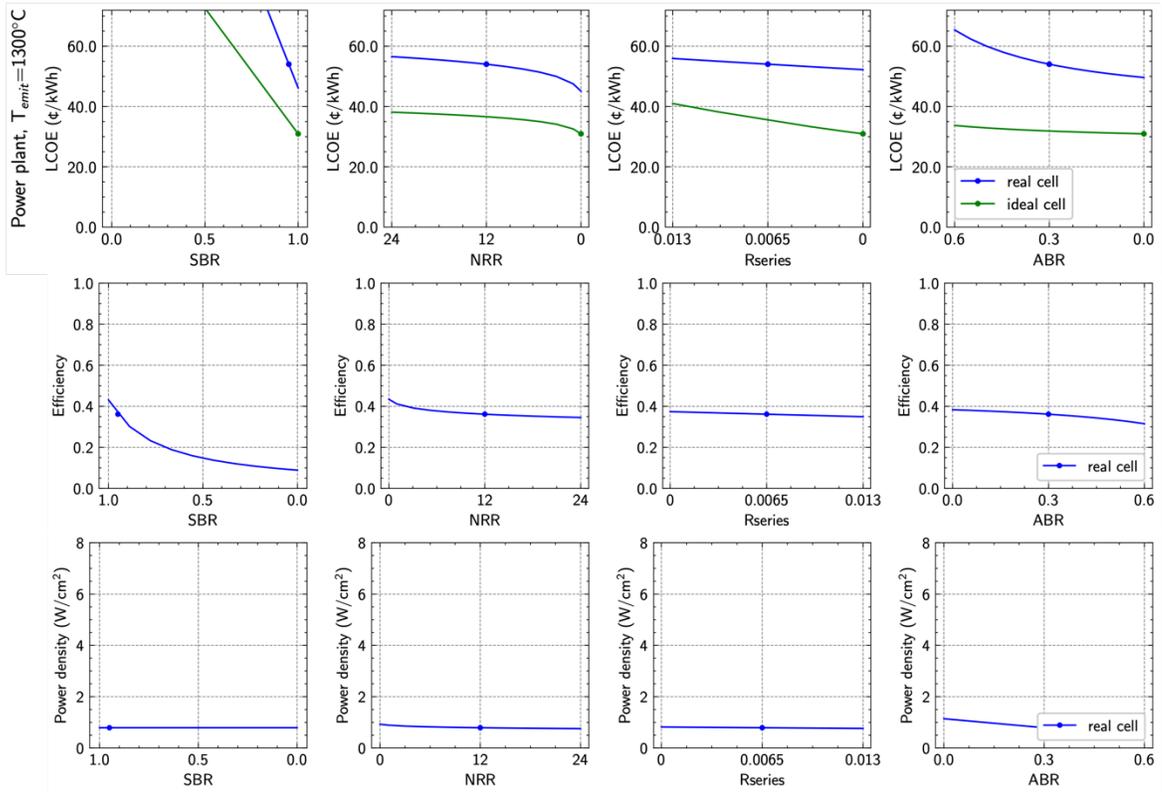
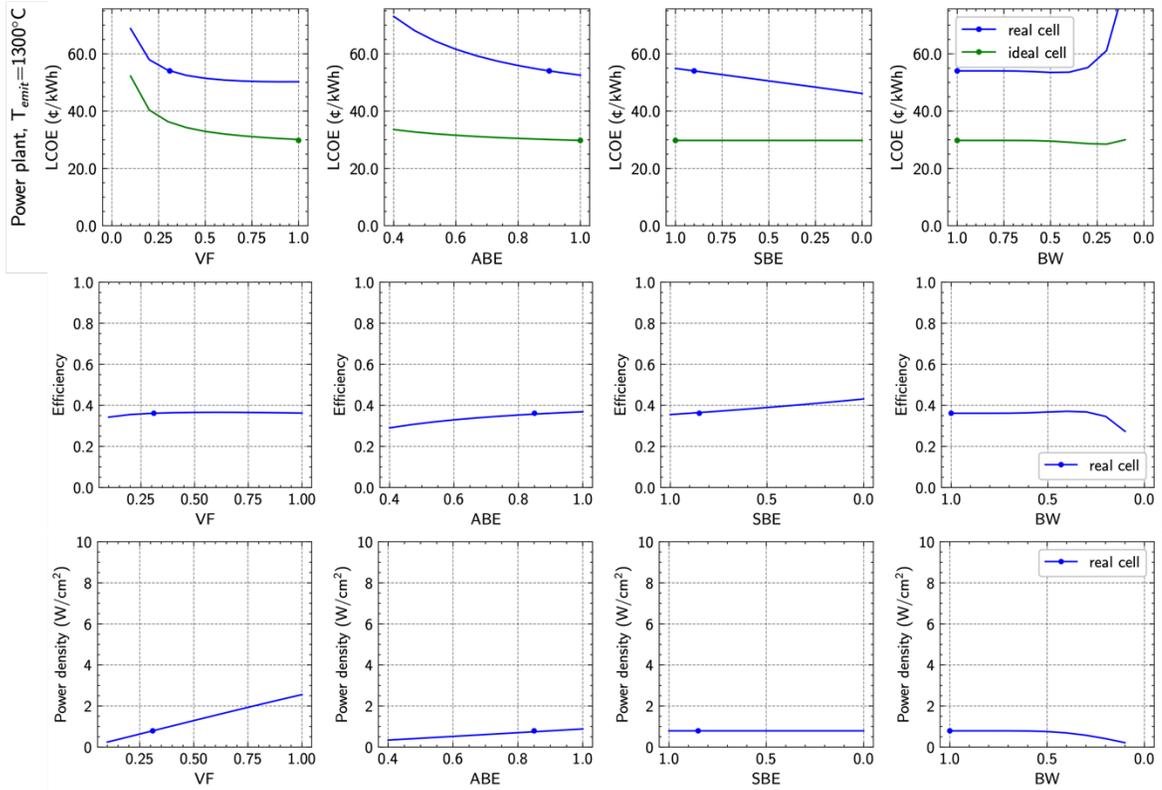

*Figure S5: LCOE, efficiency and power density for the efficiency-limited case with emitter temperature 1300C and cell bandgap 0.65eV. Model of a real cell with properties provided in Table 4 while varying a single property on each x-axis.*

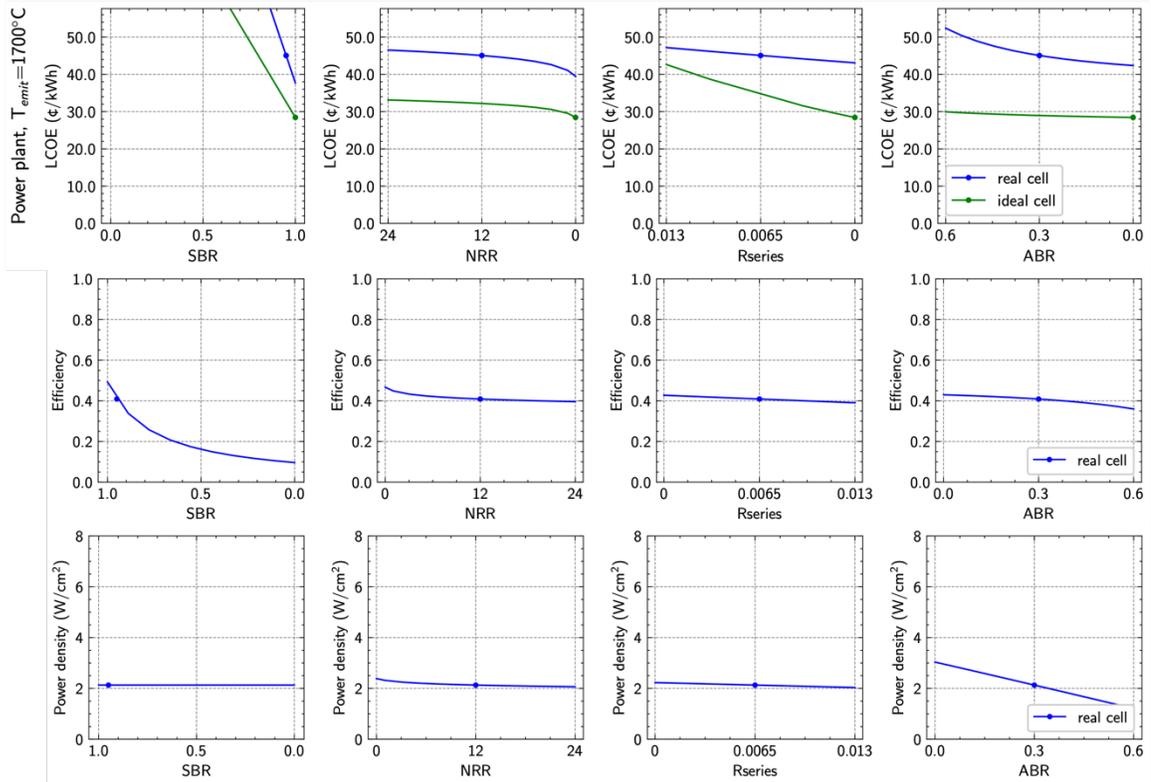
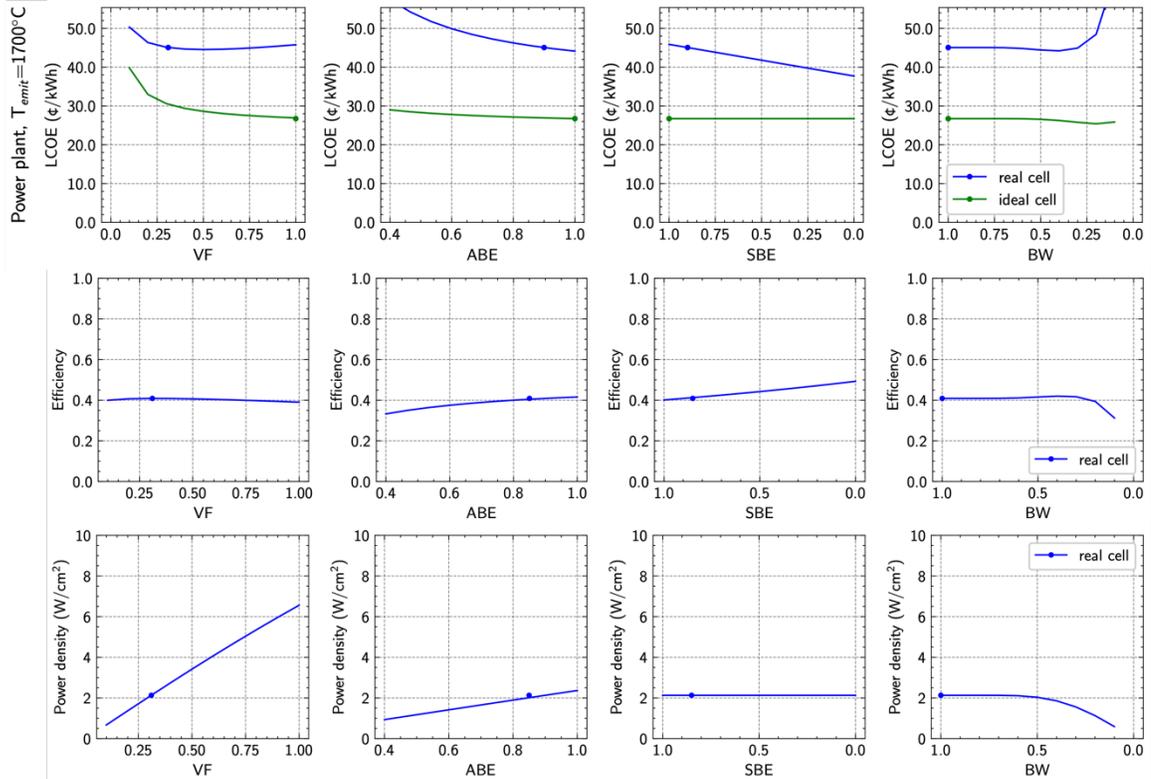

*Figure S6: LCOE, efficiency and power density for the efficiency-limited case with emitter temperature 1700C and cell bandgap 0.83eV. Model of a real cell with properties provided in Table 4 while varying a single property on each x-axis.*

# Efficiency-limited, T=2150C, BG = 1.05 eV

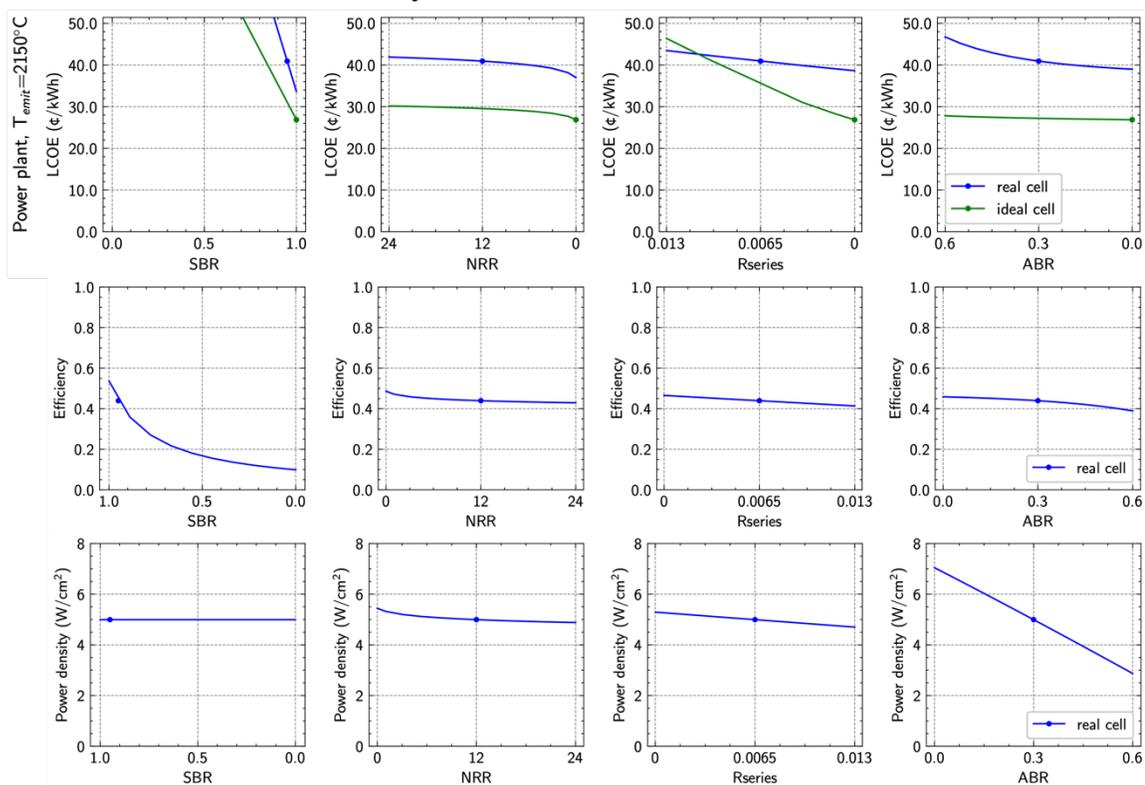

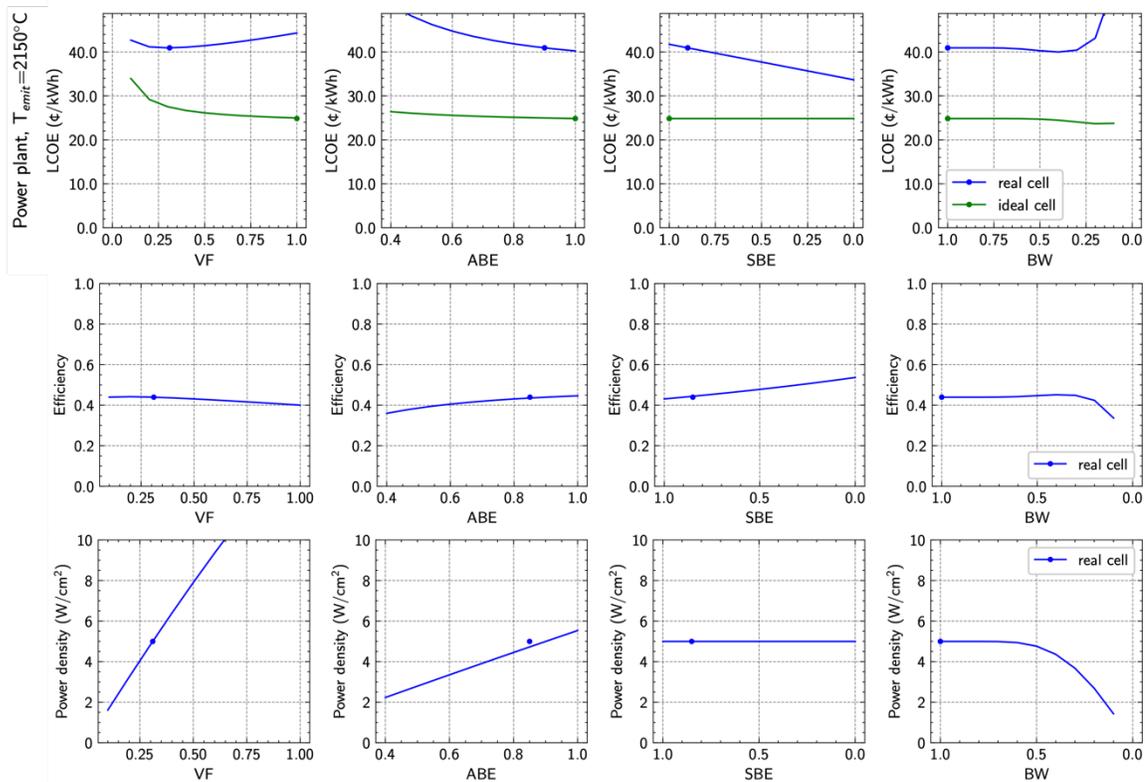

*Figure S7: LCOE, efficiency and power density for the efficiency-limited case with emitter temperature 2150C and cell bandgap 1.05eV. Model of a real cell with properties provided in Table 4 while varying a single property on each x-axis.*

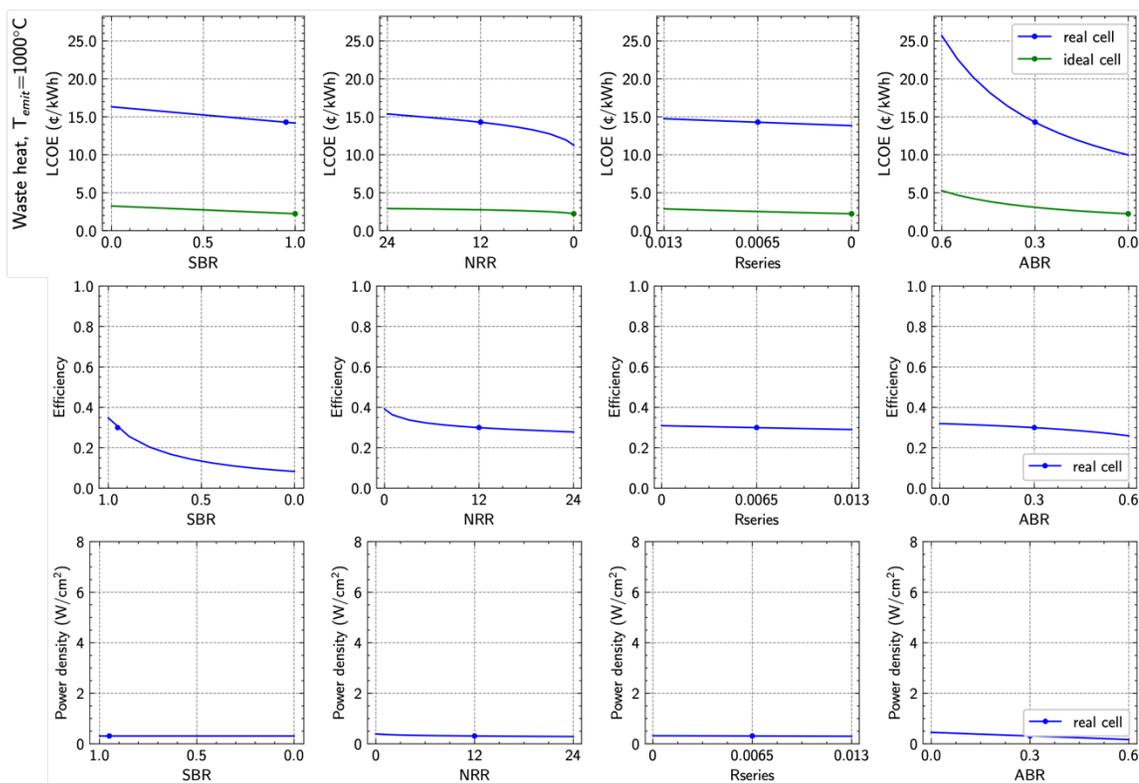
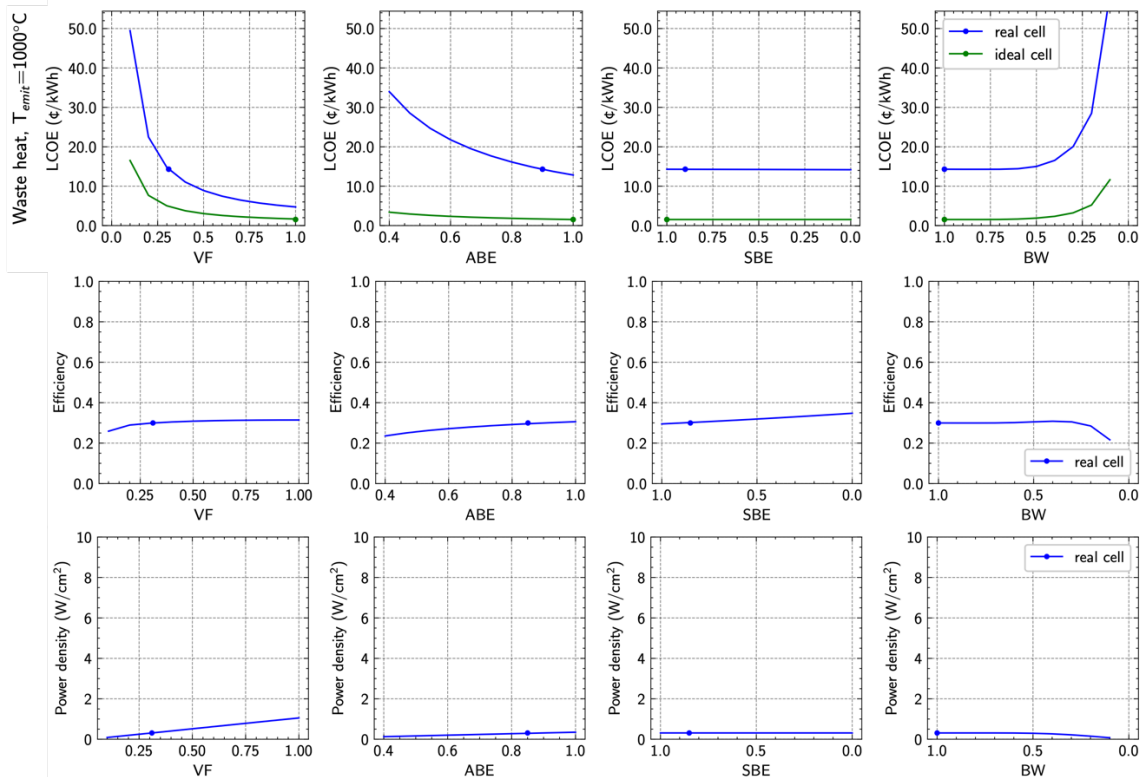

*Figure S8: LCOE, efficiency and power density for the power-limited case with emitter temperature 1000C and cell bandgap 0.50eV. Model of a real cell with properties provided in Table 4 while varying a single property on each x-axis.*

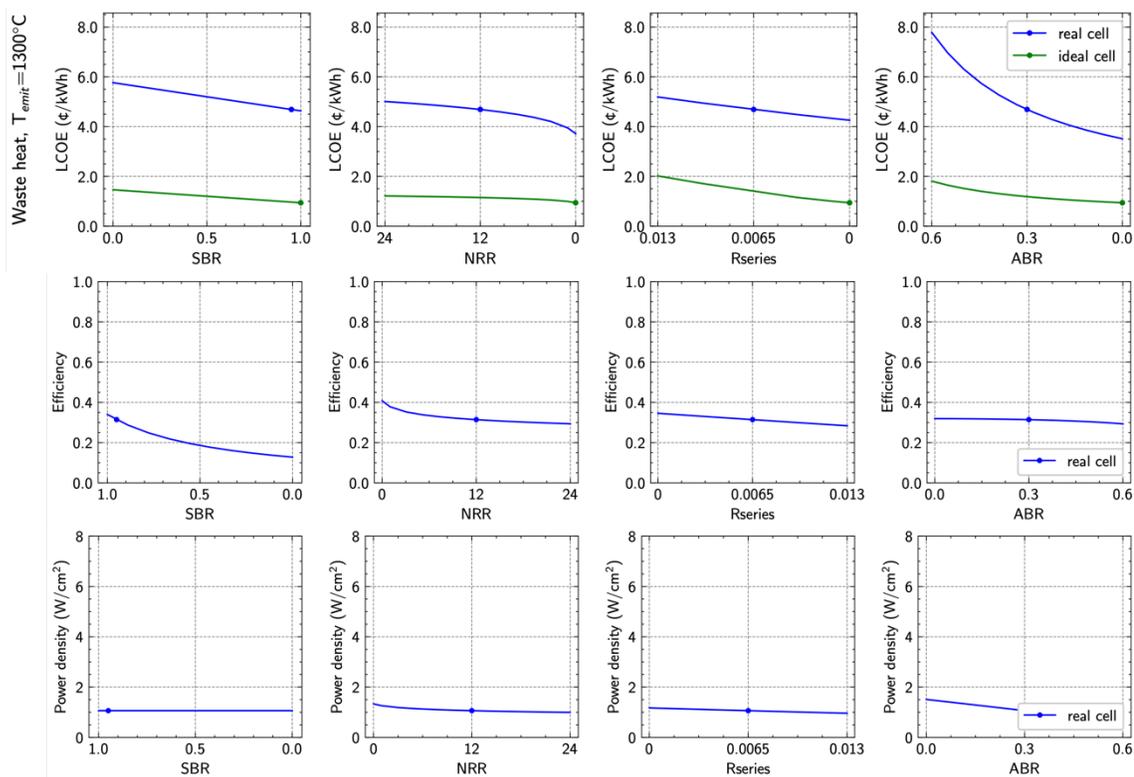
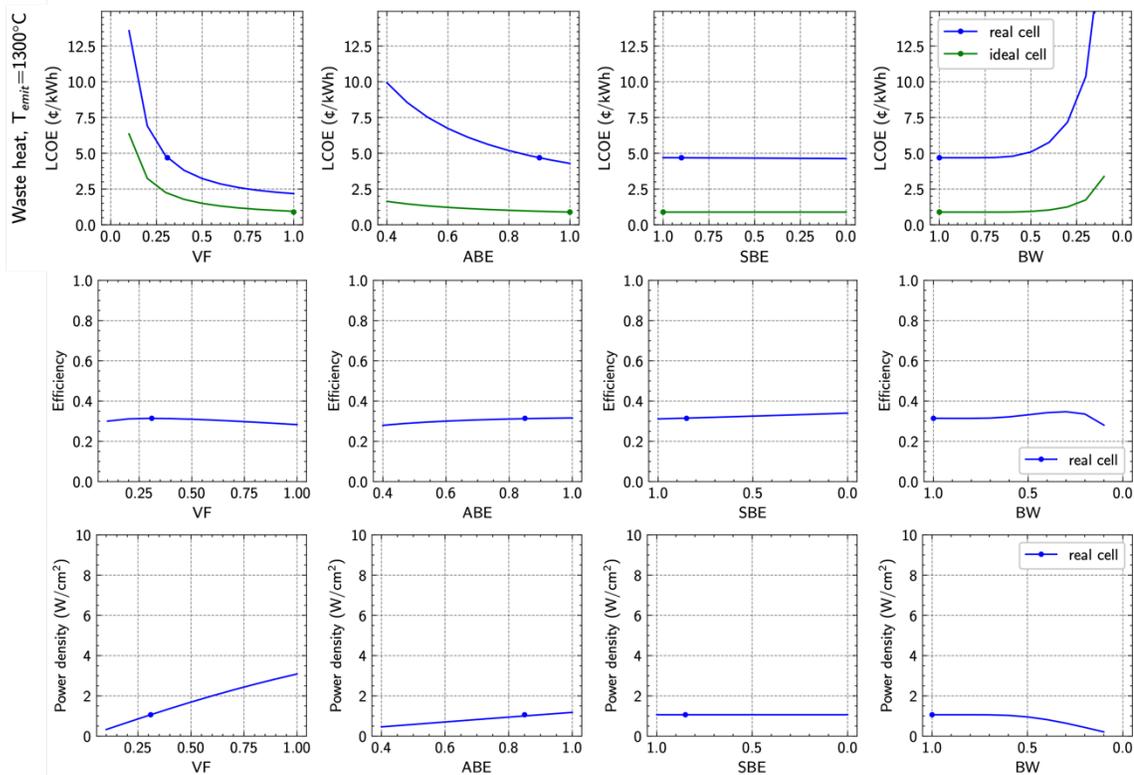

*Figure S9: LCOE, efficiency and power density for the power-limited case with emitter temperature 1300C and cell bandgap 0.50eV. Model of a real cell with properties provided in Table 4 while varying a single property on each x-axis.*

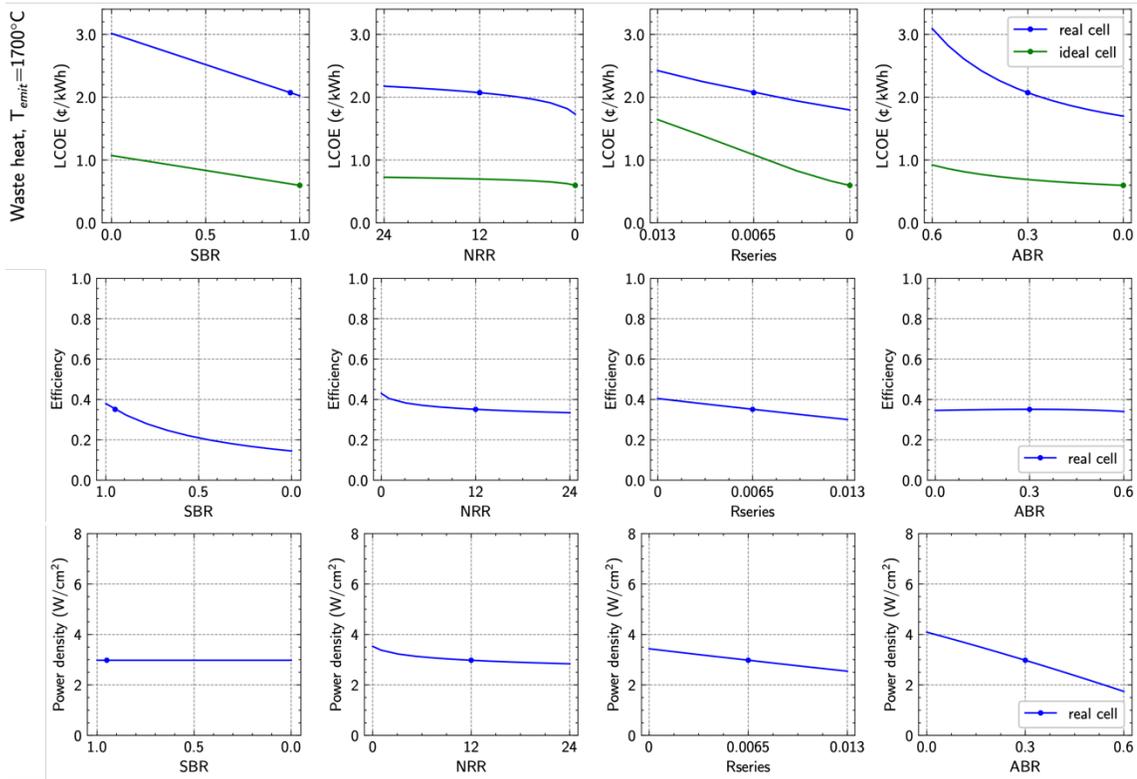
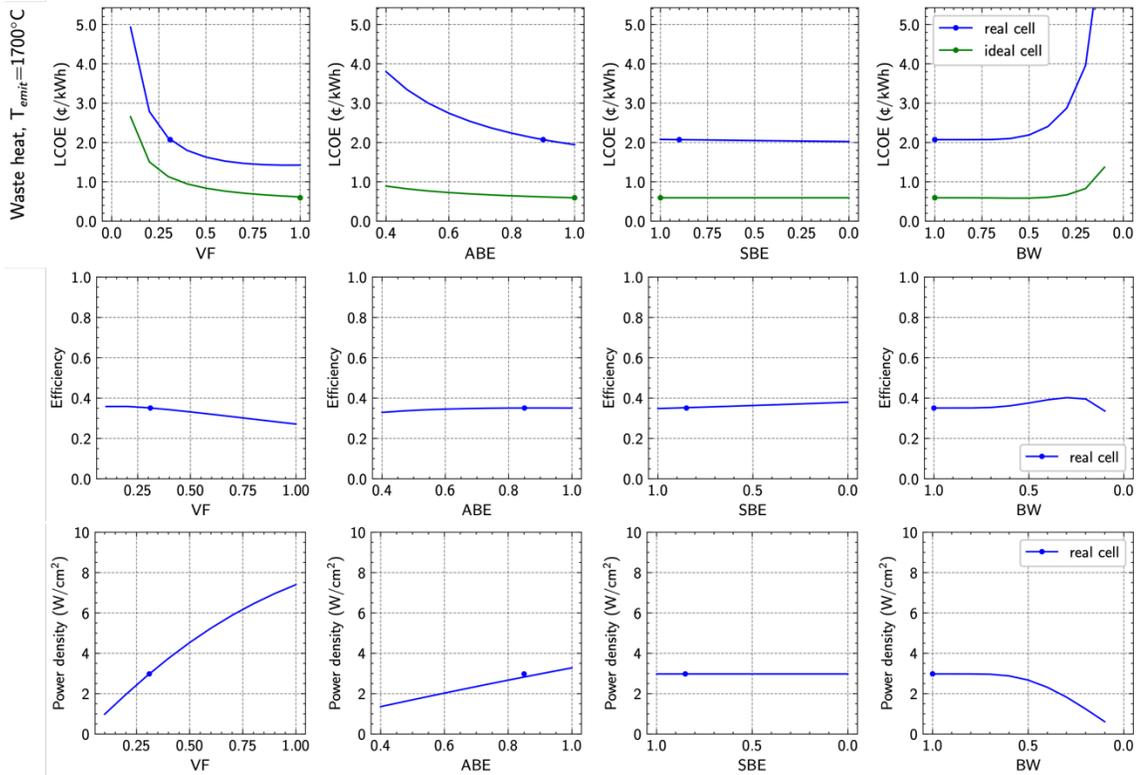

*Figure S10: LCOE, efficiency and power density for the power-limited case with emitter temperature 1700C and cell bandgap 0.63eV. Model of a real cell with properties provided in Table 4 while varying a single property on each x-axis.*

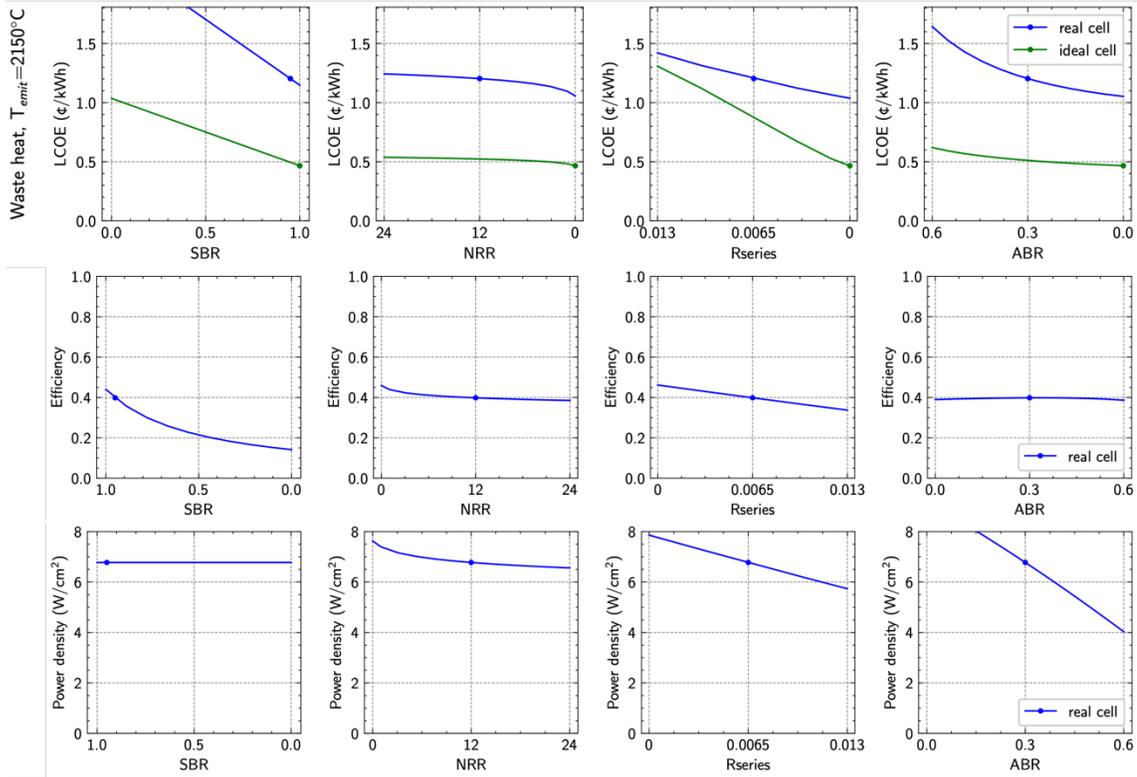
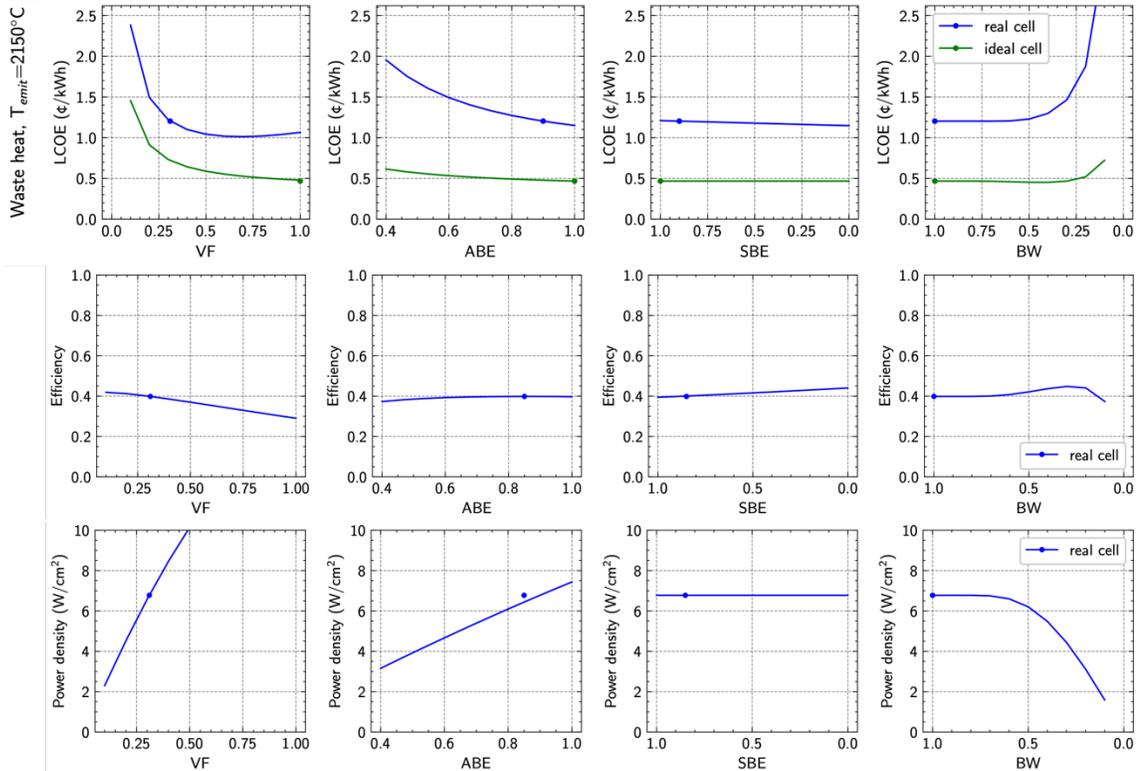

*Figure S11: LCOE, efficiency and power density for the power-limited case with emitter temperature 2150C and cell bandgap 0.83eV. Model of a real cell with properties provided in Table 4 while varying a single property on each x-axis.*

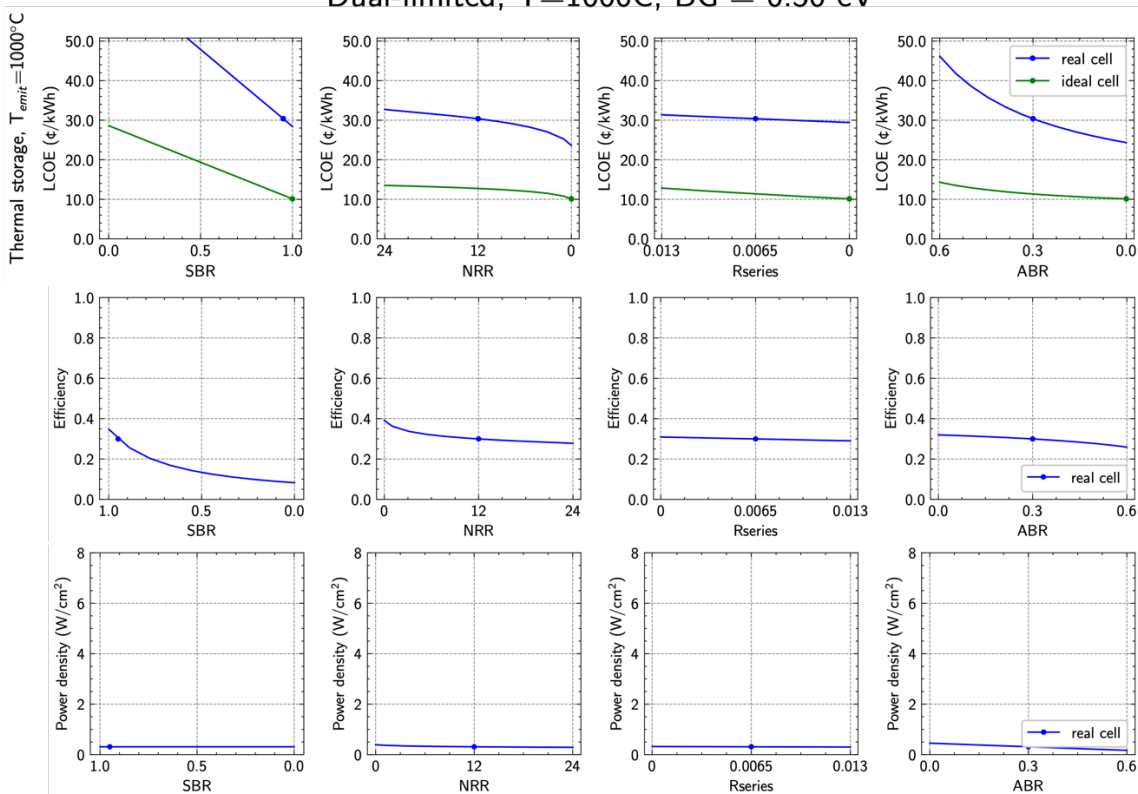
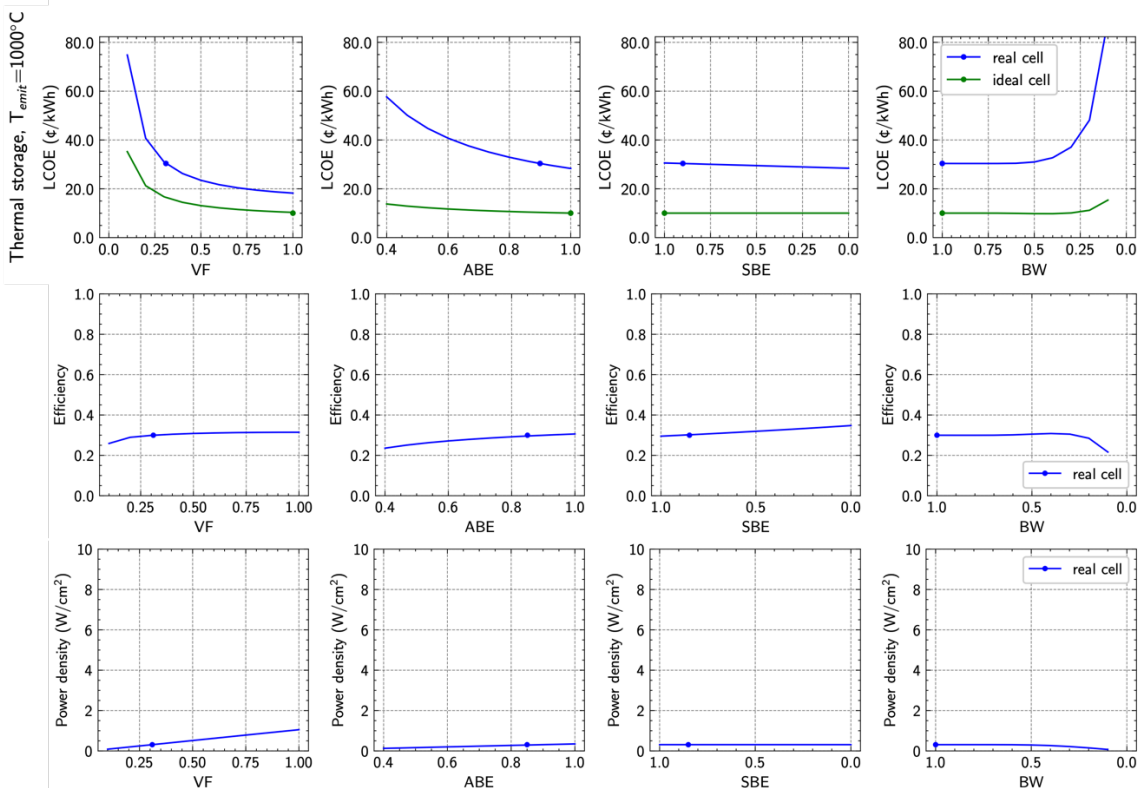

*Figure S12: LCOE, efficiency and power density for the dual-limited case with emitter temperature 1000C and cell bandgap 0.50eV. Model of a real cell with properties provided in Table 4 while varying a single property on each x-axis.*

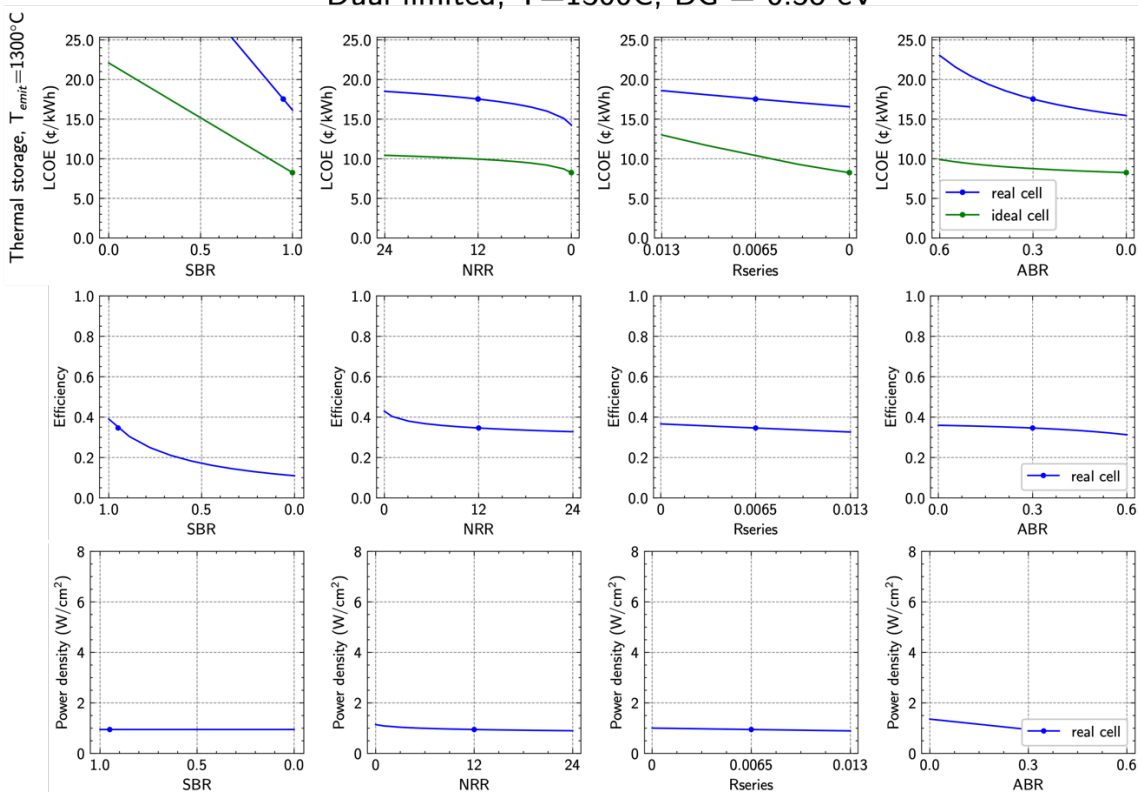
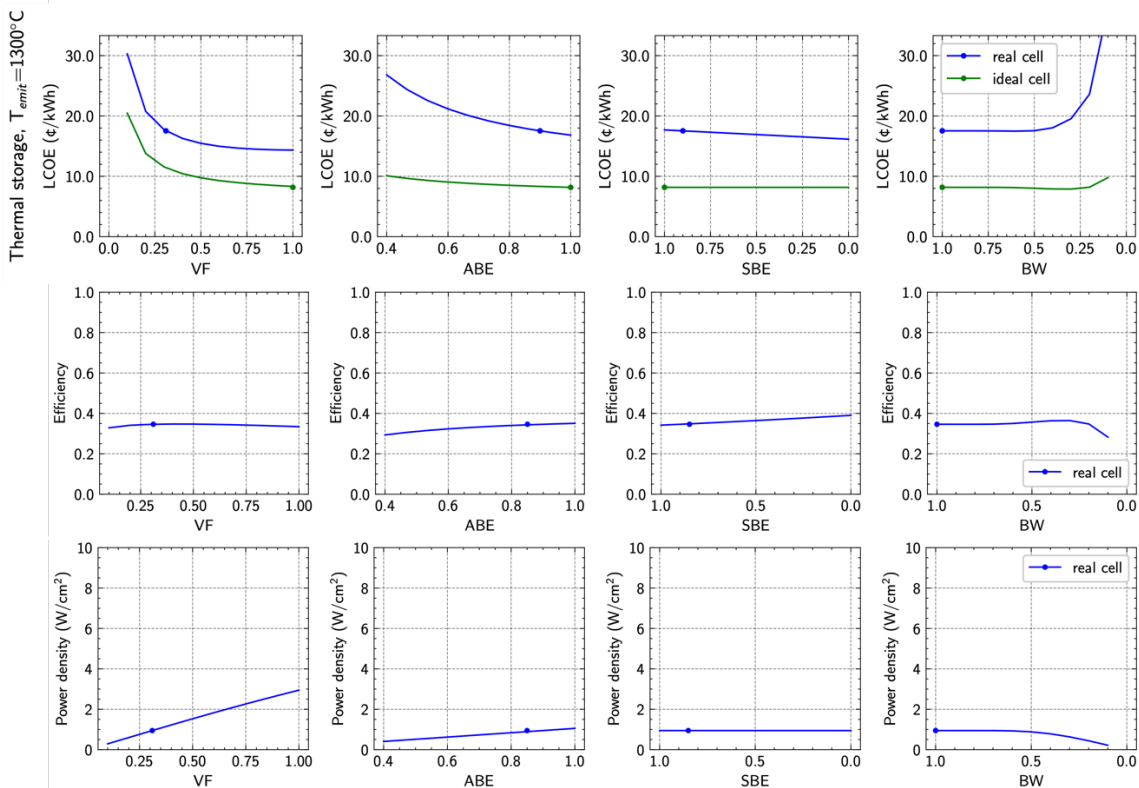

*Figure S13: LCOE, efficiency and power density for the dual-limited case with emitter temperature 1300C and cell bandgap 0.58eV. Model of a real cell with properties provided in Table 4 while varying a single property on each x-axis.*

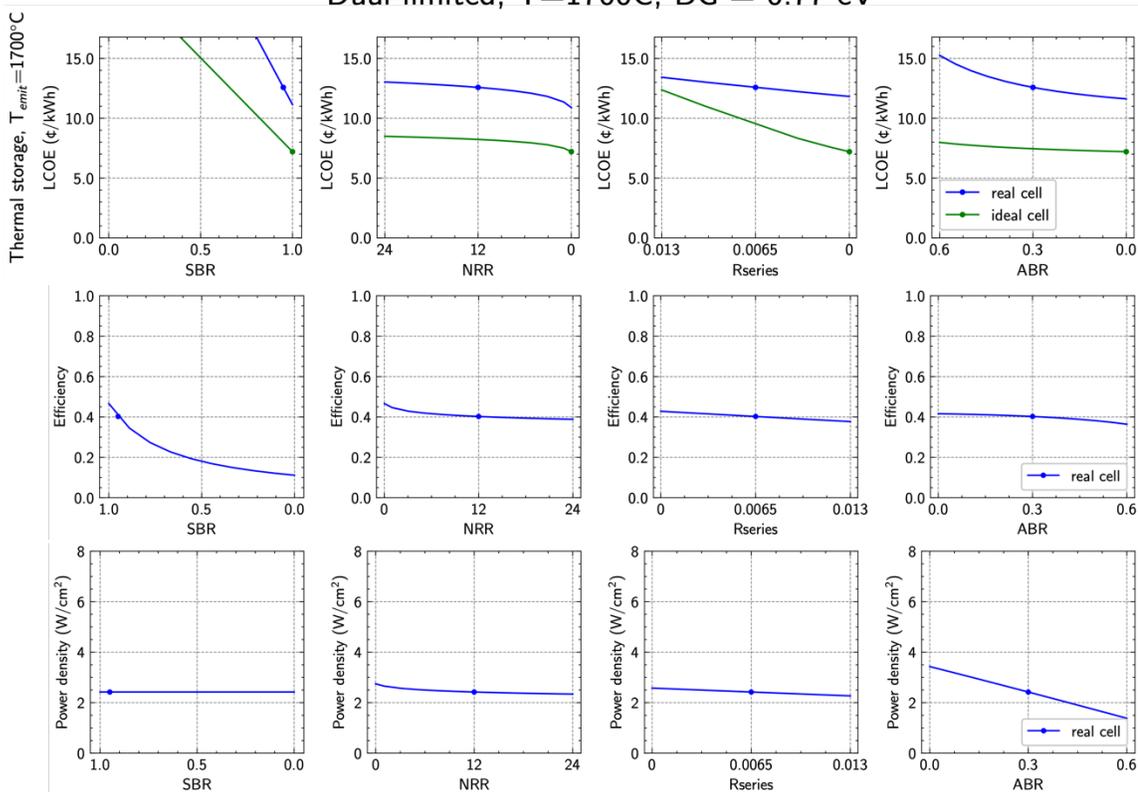
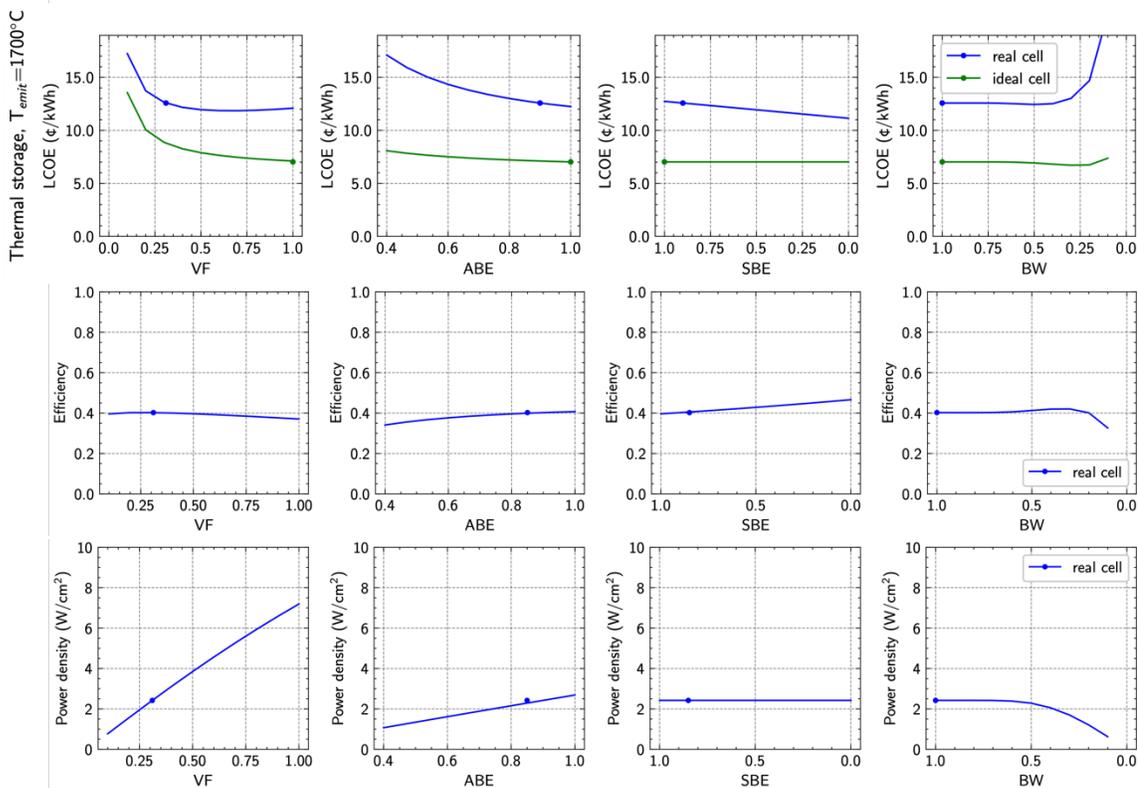

*Figure S14: LCOE, efficiency and power density for the dual-limited case with emitter temperature 1700C and cell bandgap 0.77eV. Model of a real cell with properties provided in **Table 5** while varying a single property on each x-axis.*

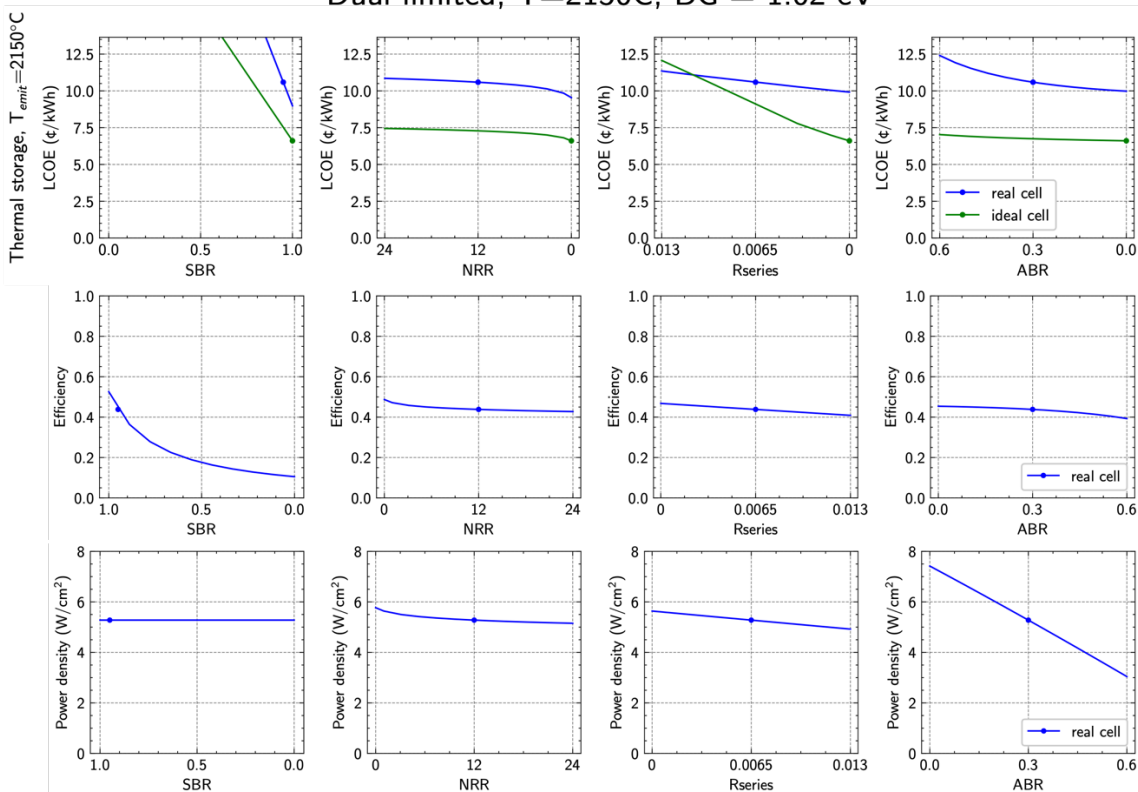
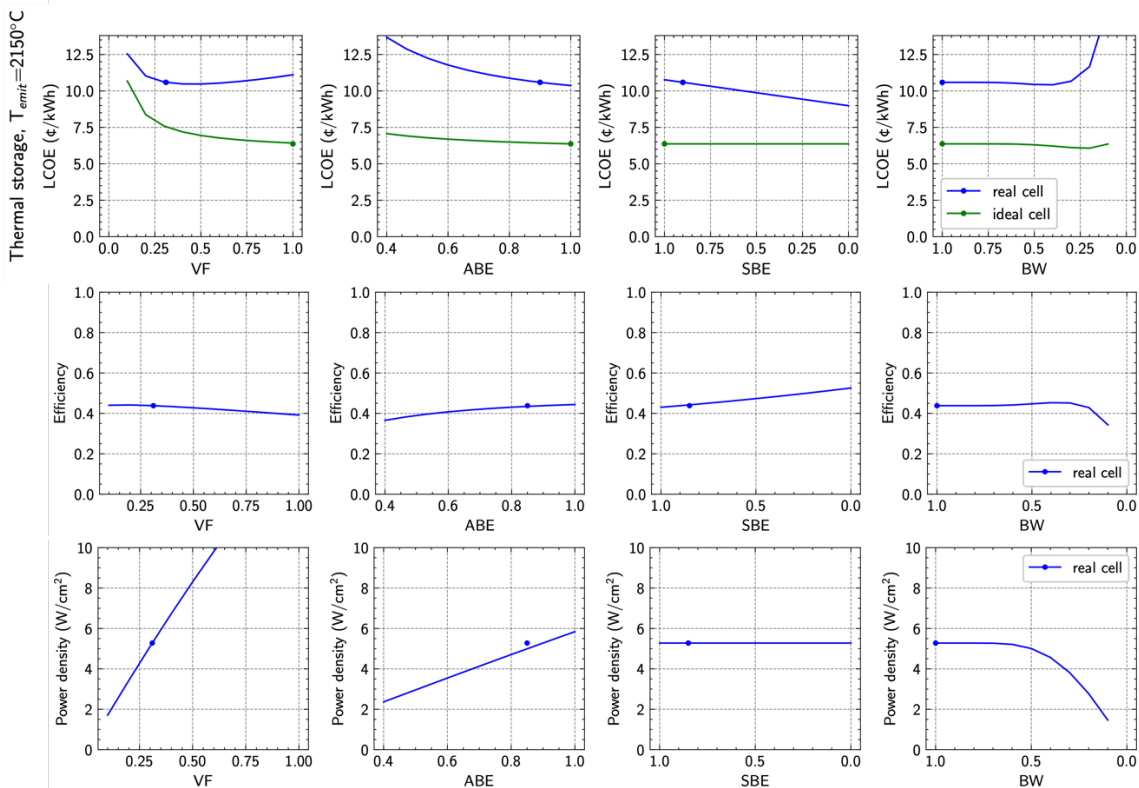

*Figure S15: LCOE, efficiency and power density for the dual-limited case with emitter temperature 2150C and cell bandgap 1.02eV. Model of a real cell with properties provided in Table 4 while varying a single property on each x-axis.*

## S8. Ranking TPV cell property improvements by importance

We propose the following methodology to rank TPV cell or emitter improvements by order of importance. First establish a base case LCOE using the properties provided in Table 4. Next, vary each individual parameter and determine which parameter allows achieving the minimum LCOE. Then set that parameter to its optimal value. Then, repeat the process for remaining variables. This allows developing a ranking of individual parameters based on their impact on LCOE. We conduct this ranking analysis for the 3 cost cases: efficiency-limited, power-limited, and dual-limited, with the results shown in Table S1. We keep the bandgap fixed to the optimal value of the real cell in Figure 5 and temperature fixed based on the application from Table 4.

*Table S1: Ranking of TPV cell and emitter improvements importance for efficiency-limited, power-limited, and dual-limited cases. Cost parameters the same as used in Figure 3(c-f).*

| Efficiency-limited case | | | |
|---|---|---|---|
| Ranking | Property | Value | LCOE ($/MWh-e) |
| - | Base case | From Table 4 | 450.57 |
| 1 | SBR | 1 | 377.11 |
| 2 | NRR | 0 | 329.1 |
| 3 | BW | 0.2 | 301.33 |
| 4 | VF | 1 | 277.53 |
| 5 | $R_{series}$ ($\Omega$ cm$^2$) | 0 | 260.04 |
| 6 | ABR | 0 | 253.54 |
| 7 | ABE | 1 | 251.89 |
| | Ideal case | From Table 4 | 250.54 |
| **Power-limited case** | | | |
| Ranking | Property | Value | LCOE ($/MWh-e) |
| - | Base case | From Table 4 | 143.05 |
| 1 | VF | 1 | 47.31 |
| 2 | ABR | 0 | 35.27 |
| 3 | NRR | 0 | 27.9 |
| 4 | $R_{series}$ ($\Omega$ cm$^2$) | 0 | 24.98 |
| 5 | ABE | 1 | 22.75 |
| 6 | SBR | 1 | 22.3 |
| 7 | BW | 0.9 | 22.3 |
| | Ideal case | From Table 4 | 15.4 |
| **Dual-limited case** | | | |
| Ranking | Property | Value | LCOE ($/MWh-e) |
| - | Base case | From Table 4 | 105.89 |
| 1 | SBR | 1 | 89.85 |
| 2 | NRR | 0 | 80.73 |
| 3 | BW | 0.3 | 75.93 |
| 4 | VF | 0.8 | 71.03 |
| 5 | $R_{series}$ ($\Omega$ cm$^2$) | 0 | 63.75 |

| 6 | ABR | 0 | 61.79 |
| 7 | ABE | 1 | 61.27 |
|   | Ideal case | From Table 4 | 59.03 |